\documentclass[11pt,a4paper]{article}

\pdfoutput=1
\usepackage{jheppub}
\usepackage{graphicx}
\usepackage{amssymb}
\usepackage{amsmath}
\usepackage{mathtools}
\usepackage{bm}
\usepackage{latexsym}
\usepackage{epsfig}
\usepackage{float}
\usepackage{textcomp}
\usepackage{color}
\usepackage{float}
\usepackage{blindtext}
\usepackage{enumitem}
\usepackage{xcolor}
\usepackage[section]{placeins}
\usepackage{graphicx}
\usepackage{subcaption}
\usepackage{tabularx}    % flexible-width columns
\usepackage{ragged2e}    % for \RaggedRight
\usepackage{multirow}
\usepackage{makecell}    % for nice multi-line headers
\usepackage{booktabs}    % nicer rules (optional)
\renewcommand{\arraystretch}{1.5}
\usepackage{float}
\usepackage{tikz} % To generate the plot from CSV
\usepackage{pgfplots}
\usepackage{pdflscape}   % for landscape pages
\usepackage{tabularx}    % flexible tables
\usepackage{multirow}    % multirow cells
\usepackage{placeins}

\usepackage{graphicx}
\pgfplotsset{compat=newest} % Allows to place the legend below the plot
\usepgfplotslibrary{units} % Allows to enter the units nicely

\usepackage{calligra}
\DeclareMathAlphabet{\mathcalligra}{T1}{calligra}{m}{n}
\DeclareFontShape{T1}{calligra}{m}{n}{<->s*[2.2]callig15}{}

\usepackage[mathscr]{euscript}
\usepackage[export]{adjustbox}
\usepackage{amsmath}
\usepackage{booktabs}
\usepackage{tabularx}

\makeatletter
\gdef\@fpheader{}
\makeatother

\def\l{\left}
\def\r{\right}
\def\d{{\rm d}}

\def\beq{\begin{equation}}
\def\eeq{\end{equation}} 
\def\be{\begin{eqnarray}}
\def\ee{\end{eqnarray}}

\definecolor{lime}{HTML}{A6CE39}
\DeclareRobustCommand{\orcidicon}{
	\begin{tikzpicture}
	\draw[lime, fill=lime] (0,0) 
	circle [radius=0.2] 
	node[white] {{\fontfamily{qag}\selectfont \tiny ID}};
	\draw[white, fill=white] (-0.0625,0.095) 
	circle [radius=0.007];
	\end{tikzpicture}
	\hspace{-2mm}
}

\def\be{\begin{equation}}
\def\ee{\end{equation}} 
\def\bea{\begin{eqnarray}}
\def\eea{\end{eqnarray}}

\def\l{\left}
\def\r{\right}
\def\d{\mathrm{d}}

\def\lsim{\mathrel{\rlap{\lower4pt\hbox{\hskip0.5pt$\sim$}}
 \raise1pt\hbox{$<$}}}         %less than or approx. symbol
\def\gsim{\mathrel{\rlap{\lower4pt\hbox{\hskip0.5pt$\sim$}}
 \raise1pt\hbox{$>$}}}         %greater than or approx. symbol

\newcommand{\ba}{\begin{aligned}}
\newcommand{\ea}{\end{aligned}}

\newcommand{\x}{\mathbf{x}}

\def\l{\left}
\def\r{\right}
\def\d{{\rm d}}

\def\be{\begin{equation}}
\def\ee{\end{equation}} 
\def\bea{\begin{eqnarray}}
\def\eea{\end{eqnarray}}

\foreach \x in {A, ..., Z}{\expandafter\xdef\csname orcid\x\endcsname{\noexpand\href{https://orcid.org/\csname orcidauthor\x\endcsname}
			{\noexpand\orcidicon}}
}

\def\l{\left}
\def\r{\right}
\def\d{{\rm d}}

\def\be{\begin{equation}}
\def\ee{\end{equation}} 
\def\bea{\begin{eqnarray}}
\def\eea{\end{eqnarray}}

\numberwithin{equation}{section}

\begin{document}

% --- Document Begins ---

\title{ \boldmath Extended mass distribution of PBHs during QCD phase transition: SGWB and mini-EMRIs}

\author[1]{Nilanjandev Bhaumik\orcidA{}}
\author[2]{Huai-Ke Guo\orcidB{}}
\author[3]{Si-Jiang Liu}

\affiliation{International Center for Theoretical Physics Asia-Pacific, \\Taiji Laboratory for Gravitational Wave Universe, \\ University of Chinese Academy of Sciences, Beijing 100190, China}
%\affiliation[b]{International Centre for Theoretical Physics Asia-Pacific, Beijing 100190, China}
%\affiliation[c]{International Centre for Theoretical Physics Asia-Pacific, Beijing 100190, China}
\emailAdd{nilanjandevbhaumik@gmail.com}
\emailAdd{guohuaike@ucas.ac.cn}
\emailAdd{liusijiang22@mails.ucas.ac.cn}
\date{\today}

\abstract{
Primordial black holes (PBHs) are one of the most important tracers of cosmic history. In this work, we investigate the formation of PBHs around the time of the QCD phase transition from a broadly peaked inflationary scalar power spectrum, which naturally produces an extended PBH mass function. This scenario yields two distinct stochastic gravitational wave backgrounds (SGWB): (i) scalar-induced, second-order tensor perturbations generated at PBH formation, and (ii) a merger-driven SGWB from the subsequent PBH binary population. Using Bayesian analysis, we examine both SGWB channels with the data from the NANOGrav 15-year dataset and the first three observing runs of LVK. We also forecast continuous-wave signals from mini extreme mass ratio inspirals (mini-EMRIs) for direct comparison with NANOGrav and LVK constraints. Our parameter scans identify regions of the parameter space where the combined SGWB is detectable in future ground-based and space-based detectors. A broad PBH mass distribution naturally gives rise to mini-EMRIs, which future ground-based observatories, such as LVK A+, ET, and CE, can detect. For a large part of the PBH parameter space, the SGWB of astrophysical origin masks the primordial SGWB in the frequency band of ground-based detectors. Thus, for extended PBH mass distributions, we find that the detection of mini-EMRIs is a more robust channel for probing the PBH parameter space than the corresponding SGWB.
}
%\end{abstract}
%\begin{document}
\maketitle
\flushbottom
\tableofcontents

\section{Introduction}
\label{sec:introduction}

The direct detection of gravitational waves in LIGO-Virgo-KAGRA (LVK) ~\cite{LIGOScientific:2016aoc} has started a new era of probing our universe with gravitational wave (GW) astronomy. LVK has already revealed a rich population of binary black holes (BH) and neutron stars(NS) systems ~\cite{LIGOScientific:2018mvr, LIGOScientific:2018jsj, LIGOScientific:2020ibl, KAGRA:2021vkt, KAGRA:2021duu}. Many of these discoveries of massive BHs have challenged conventional models of stellar evolution to form black holes, as standard astrophysical processes such as the gravitational collapse of massive stars are unlikely to produce very massive black holes ~\cite{Fryer_2001}. This intensified interest in the Primordial black holes (PBHs) \cite{DeLuca:2025fln, Yuan:2025avq}. First proposed by Hawking and Carr~\cite{Hawking:1971ei, Carr:1974nx}, PBHs are a class of black holes that formed in the very early Universe from the collapse of large-amplitude density perturbations. Unlike astrophysical black holes (ABHs), which form due to the stellar collapse, the formation of PBHs is not subject to the constraints of nuclear processes, allowing them to span a vast and continuous range of masses—from the Planck scale to supermassive objects~\cite{Escriva:2022duf, Carr:2020xqk, Carr:2020xqk1, Carr:2009jm}. On the other hand, the channels of stellar collapse are not expected to produce sub-solar mass black holes. Thus, the detection of sub-solar mass black holes through GW observatories would be a smoking gun test for the existence of PBHs~\cite{Carr:2020xqk, PhysRevLett.126.141105, PhysRevLett.127.151101, 10.1093/mnras/stad588}. The formation of PBHs in early Universe also accompany the amplification in the second order tensor perturbation or the stochastic gravitational wave background (SGWB)\cite{Ananda:2006af, Baumann:2007zm, Saito:2008jc, Kohri:2018awv, Espinosa:2018eve, Bhaumik:2020dor, Maity:2024odg, Zeng:2025cer} which has also been considered as a potential candidate to explain the recent and first of its kind detection of SGWB in Pulsar Timing Array (PTA) collaborations, in the North American Nanohertz Observatory for Gravitational Waves (NANOGrav)~\cite{NANOGrav:2023gor}, the European Pulsar Timing Array (EPTA)~\cite{EPTA:2023sfo, EPTA:2023fyk}, 
the Parkes Pulsar Timing Array (PPTA)~\cite{Zic:2023gta, Reardon:2023gzh}, and the Chinese Pulsar Timing Array (CPTA)~\cite{Xu:2023wog}. The subsequent Bayesian analysis revealed that the NANOGrav 15-year data are more consistent with SGWB of primordial origins than the astrophysical explanation originating from the merging of supermassive BH binaries (SMBHBs)~\cite{NANOGrav:2023hvm,
 NANOGrav:2023gor}. 

 However, the importance of PBHs in the context of cosmology is not so much due to their prospects in explaining the GW observations, but rather to understanding the broader cosmic history, as PBHs have profound implications for different stages of cosmic evolution, starting from inflation. They have the potential to solve or explain many of the inconsistencies of standard cosmology \cite{Carr:2019kxo}. This capability underlies renewed interest in PBHs as dark-matter candidates~\cite{Carr:2021bzv, Bird:2016dcv, Green:2020jor,  Khlopov:2008qy}. Moreover, PBHs act as a tracer of large amplifications in the small-scale scalar power spectrum and thereby the inflationary models leading to such amplifications\cite{Garcia-Bellido:2017mdw, Ezquiaga:2017fvi, Bhaumik:2019tvl, Bhaumik:2020dor, Ragavendra:2020sop, Braglia:2020eai, Balaji:2022zur, Kumar:2025jfi, Maiti:2025ijr}: while measurements of the cosmic microwave background and large-scale structure tightly constrain the spectrum over the first few e-folds of inflation, the subsequent $\sim$40--50 e-folds remain largely unconstrained observationally \cite{Guth:1980zm, Linde:1981mu, Planck:2018jri}. Enhancements of scalar perturbations on these small scales—for example, from features in the inflationary potential or non-standard reheating dynamics—can therefore seed substantial PBH production \cite{Garcia-Bellido:2017mdw, Ezquiaga:2017fvi}. Any viable PBH-formation scenario must satisfy two key observational checks: stringent upper limits on PBH abundance from dark matter and astrophysical constraints~\cite{Carr:2017jsz, Green:2020jor}, and the unavoidable generation of a scalar-induced stochastic gravitational-wave background (SGWB) at second order in perturbation theory sourced by the same enhanced scalar modes~\cite{Ananda:2006af, Baumann:2007zm, Saito:2008jc, Kohri:2018awv}.

 Much of the literature on the context of PBHs focuses on the near monochromatic mass spectrum of PBHs, often originating from a sharply peaked inflationary scalar power spectrum~\cite{Carr:2017jsz, Bhaumik:2019tvl, Bhaumik:2020dor, Bhaumik:2022pil, Bhaumik:2022zdd, Bhaumik:2023zcz}. However, the extended mass distribution of PBHs can occur naturally and leave unique observational footprints \cite{Carr:2017jsz, Laha:2018zav}. We investigate PBH formation in a broader mass distribution from a broadly peaked power spectrum~\cite{DeLuca:2020ioi, Braglia:2021wwa} around the epoch of the QCD phase transition~\cite{Jedamzik:1996mr, Byrnes:2018clq, Musco:2020jjb}. The softening of the Universe's equation of state during this period lowers the threshold for gravitational collapse, thereby enhancing PBH production. Assuming a flat amplification in the inflationary power spectrum naturally leads to an extended mass function of PBHs, where the broadness of the spectrum directly depends on the broadness of the scalar power spectra peak, and it can reasonably span from sub-solar to several solar masses.  

This gives rise to a rich, multi-channel gravitational-wave phenomenology, where on the SGWB channel we get two very different origin of SGWB, namely the second-order induced SGWB ~\cite{Domenech:2021ztg} in the low frequencies ( nHz band in our case, consistent with the recent PTA findings) and the SGWB coming from the mergers of this broad population of PBHs, which typically peaks at relatively higher frequencies \cite{zphk-3ld9, Andres-Carcasona:2024wqk}, placing it within the sensitivity bands of ground-based detectors like LVK~\cite{LIGOScientific:2014pky, LIGOScientific:2016fpe}, the Einstein Telescope (ET)~\cite{Punturo:2010zz, Hild:2010id}, and Cosmic Explorer (CE)~\cite{Evans:2021gyd}, and space-based detectors like  Laser Interferometer Space Antenna(LISA)~\cite{Bartolo:2016ami, Auclair:2022lcg}, Atom Interferometer Observatory and Network (AION)~\cite{Badurina:2021rgt, Badurina:2019hst, Graham:2016plp, Graham:2017pmn} and Taiji~\cite{Luo:2021qji}. Finally, the coexistence of solar-mass and sub-solar-mass PBHs within a single population allows for the formation of detectable merger events from mini extreme mass ratio inspirals (mini-EMRI) ~\cite{Guo:2022sdd}, where both components are black holes of primordial origin.

The mini-EMRI systems are a particularly novel probe, differing from conventional EMRIs (which involve stellar-mass objects orbiting supermassive black holes) in both their formation channel and their detectability~\cite{Amaro-Seoane:2007osp, Amaro-Seoane:2012lgq}. While the constituents of standard EMRIs can form through dynamical capture in galactic centers~\cite{Pan:2021ksp} or from primordial origins \cite{Huang:2024qvz}, the lower mass constituent of the mini-EMRI must be a compact object of subsolar mass. In this case, it can not be stellar black holes. We can expect mini-EMRI systems to form in the early universe through PBH clustering or dynamics within dense primordial halos~\cite{Raidal:2018bbj, DeLuca:2020bjf, Kavanagh:2018ggo}. While EMRIs with supermassive primary PBHs are tightly constrained by Cosmic Microwave Background (CMB) accretion and $\mu$-distortion limits from Far Infrared Absolute Spectrophotometer (FIRAS)~\cite{1994ApJ...420..439M, 2012ApJ...758...76C, 2014PhRvL.113f1301J, Carr:2020gox}, mini-EMRIs composed of smaller PBHs are comparatively less affected, though still subject to microlensing bounds from surveys such as EROS~\cite{EROS-2:2006ryy, Carr:2017jsz}. The significantly smaller mass of mini-EMRI systems results in higher gravitational-wave frequencies observable in current and future ground-based observatories, such as LVK, ET, and CE, making them an ideal channel to probe or constrain the extended mass distribution of PBHs around solar masses.

In this work, we begin by performing a Bayesian analysis of the second-order induced SGWB from our PBH-forming model for NANOGrav 15-year data, identifying the parameter space consistent with the NANOGrav 15-year dataset. We perform a similar analysis using the LVK O1-O3 datasets with merger-induced SGWB, which enables us to rule out a significant portion of our parameter space. 
 Beyond current constraints, we also estimate the future discovery potential of the combined SGWB signal from our model for many ground and space-based detectors by conducting detailed parameter space scans. We find that for a range of viable parameters where the second-order induced SGWB from this scenario can explain NANOGrav 15-year data, the SGWB from PBH mergers in that parameter region can be directly probed by future GW observatories like LISA, AION, Taiji, LVK A+, ET, and CE with a signal-to-noise ratio (SNR) $>$ 10. A similar parameter space scan for the rate of detectable mini-EMRI events reveals that, for constraining or detecting a broader PBH mass fraction (characterized by a broad inflationary scalar power spectrum peak), the detection of mini-EMRI events can be a more effective channel than the corresponding stochastic background.

The structure of the paper is as follows. In \textbf{Section~\ref{sec:framework}}, we establish our PBH formation and merger estimation framework, detailing the inflationary power spectrum, the formalism for calculating the PBH mass function during the QCD phase transition, and the semi-analytical model used to determine PBH merger rates. In \textbf{Section~\ref{sec:sgwb_probes}}, we present our primary observational analysis, focusing on the two distinct SGWB channels predicted by the model. We compare the combined SGWB with the NANOGrav 15-year dataset, LVK O1-O3 data, and projected sensitivities of future ground-based and space-based detectors to analyze the parameter space. In \textbf{Section~\ref{sec:emri_probes}}, we explore the prospects for detecting mini-EMRIs, demonstrating how rates of resolvable events serve as a powerful and complementary probe of this scenario. Finally, in \textbf{Section~\ref{sec:conclusions}}, we summarize our findings and discuss the future multi-messenger outlook.
\section{PBH formation and mergers}
\label{sec:framework}

In this section, we lay the theoretical groundwork for our analysis. We begin by defining the primordial power spectrum that sources PBH formation, then detail the formalism used to calculate the PBH mass function, including the crucial effects of the QCD phase transition, and finally describe the model for calculating PBH binary merger rates.

\subsection{PBH Formation from an Enhanced Primordial Spectrum}

\subsubsection{Inflationary curvature power spectrum}
Inflation, the earliest exponentially expanding phase of our Universe, resolves the horizon and flatness problems of standard Big Bang cosmology and generates quantum perturbations to explain the CMB temperature and polarization anisotropies optimally~\cite{Planck:2018jri}.
Recently, inflationary models have been studied to obtain additional amplification in scalar perturbations on smaller scales, which can form PBHs when these amplified perturbation modes re-enter the horizon at a later stage. The  PBH-formation requires the perturbation modes to be amplified over a certain threshold value of order unity, and some specific inflationary models have been proposed to efficiently achieve such an implication without disturbing the CMB scale perturbations. In this work, we focus on a broadly peaked inflationary scalar power spectrum~\cite{Ballesteros:2018swv, Germani:2018jgr, MoradinezhadDizgah:2019wjf, DeLuca:2020ioi, DeLuca:2020agl, Franciolini:2022tfm, Franciolini:2022pav}, which is the simplest possible model to form an extended mass distribution of PBHs,
\begin{align}
%\label{eq:PR}
{\mathcal P}_{\mathcal R}(k) &=A_s \left(\frac{k}{k_{\rm p}}\right)^{n_s-1} +A_0 
    \left ( \frac{k}{k_s}\right )^{n_0-1}
    \Theta(k-k_s)
    \Theta(k_e-k) \, .
    \label{inflPS}
\end{align}
Here $A_s$ and $n_s$ are the amplitude and scalar index of the power spectrum at the CMB pivot scale of $k_p=0.05\,\mathrm{Mpc}^{-1}$  as suggested by the Planck data~\cite{Planck:2018jri}. $k_e$ and $k_s$ denote the amplified wave number range, $A_0$ is the height of the peak, $n_0$ is the tilt of the top of the amplified part, which would be $n_0=1$ for exactly flat settings. 
We shall choose appropriate values of $k_s$ and $k_e$ which lead to PBH formation around the era of the QCD phase transition. One interesting aspect of PBH formation during phase transition is the shift in the critical threshold value for PBH formation, as PBH collapse depends on the equation of state of the Universe when perturbation modes re-enter the horizon. The softening of the equation of state during the QCD phase transition  \cite{Borsanyi:2016ksw, Bhattacharya:2014ara} can therefore lead to a relatively more abundant production of PBHs at the time of QCD phase transition \cite {Chapline:1975tn, Jedamzik:1996mr, Jedamzik:1998hc, Jedamzik:1999am, Byrnes:2018clq, Carr:2019kxo, DeLuca:2020agl, Juan:2022mir, Escriva:2022bwe, Musco:2023dak, Jedamzik:2024wtq, Iovino:2024tyg, Pritchard:2024vix}.

\subsubsection{The sound speed and equation of state during QCD phase transition}
\begin{figure}[t!]
\begin{center}
\includegraphics[scale=0.46]{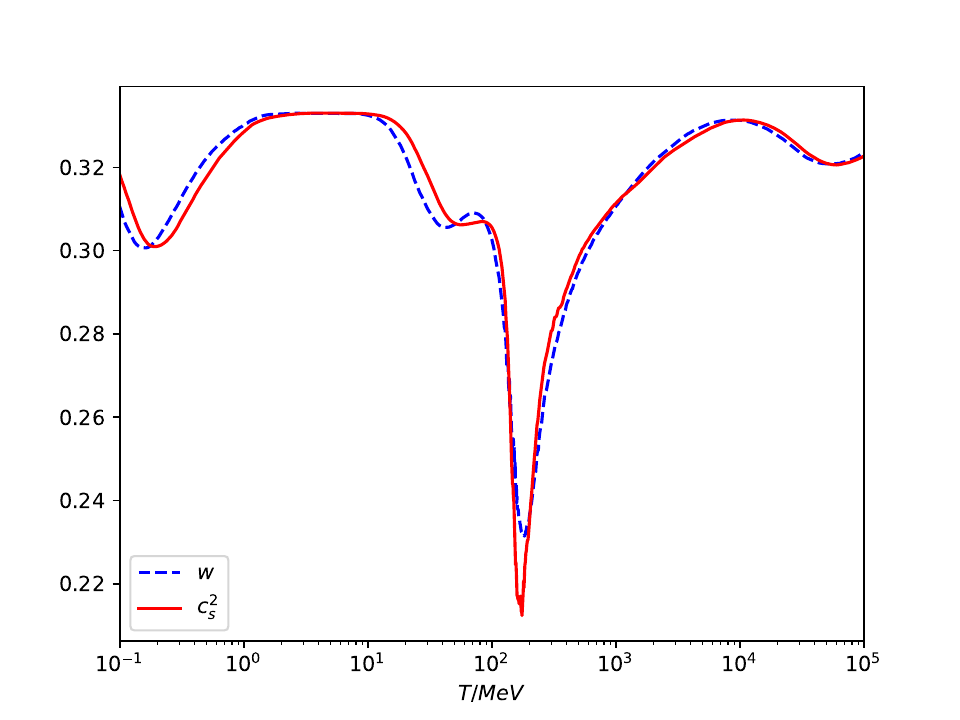}
\includegraphics[scale=0.46]{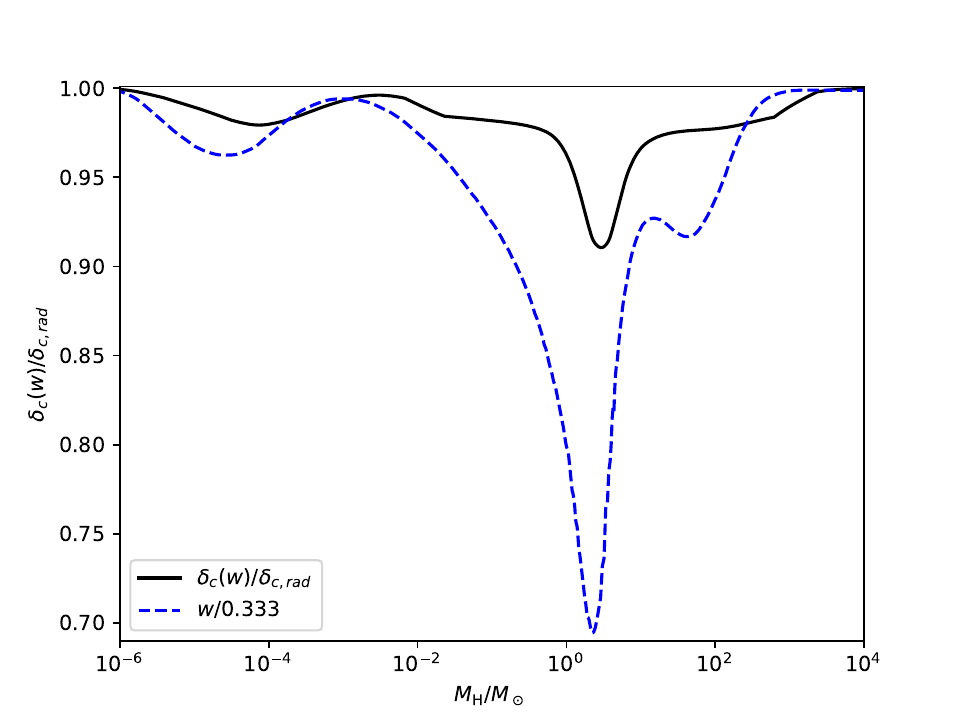}
\vskip 6pt
\caption{ \textbf{Left panel:} Here, we plot the equation of state as a function of temperature, $T$  \cite{Borsanyi:2016ksw}. \textbf{Right panel:} We plot the resultant deviation in the critical density contrast from the QCD phase transition, as obtained in \cite{Musco:2023dak}.}
\label{eosdeltac}
\end{center}
\end{figure}

The particle degrees of freedom change with cosmic time, as temperature $T$ decreases and lighter particles combine to form heavier particles. This also leads to a change in the ratio between the pressure, $p$, and the energy density of the Universe $\rho$,
\begin{equation}
w(T) \equiv \frac{p}{\rho} = \frac{4g_{*,s}(T)}{3g_{*}(T)} - 1 \,.
\end{equation}
Here $g_{*}(T)$ and $g_{*,s}(T)$ refer to the effective number of relativistic degrees of freedom for energy and entropy, 
\begin{equation}
   g_{*}(T) = \frac{30\rho}{\pi^2T^4} 
    \quad \textrm{and} \quad 
   g_{*,s}(T) = \frac{45s}{2\pi^2T^3}\,.
\end{equation}
where $s$ defines the entropy density of the Universe. We also define equation of state $w (T)$ as to connect $p$ and $\rho$,
\begin{equation}
    p=sT-\rho=w(T)\rho\,.
\end{equation}
In the left panel of Fig.~\ref{eosdeltac} where we plot the varying equation of state as a function of fluid temperature $T$, we can see that during the confinement of quarks into hadrons there is a prominent dip in the value of the equation of state $w (T)$, which shall also lead to a dip in sound speed of our universe $c_s^2\equiv \partial p/\partial \rho$ during the QCD phase transition. These results are obtained from lattice QCD simulations~\cite{Hindmarsh:2005ix, Borsanyi:2013bia}.  

\subsubsection{Estimating PBH Abundance}
To estimate the PBH abundance, we employ the Peaks Theory formalism~\cite{Bardeen:1985tr} and use the PBH collapse threshold estimated around the time of QCD phase transition \cite{Musco:2023dak}. We do not consider any non-Gaussian contribution in the inflationary scalar perturbation and use the inflationary scalar power spectrum of Eqn \eqref{inflPS} to calculate the mass fraction of the PBHs.

\paragraph{The Compaction Function and the shape parameter}
In the comoving uniform-density gauge, the perturbed Friedmann-Lemaître-Robertson-Walker (FLRW) metric  is given by,
\begin{equation}
    ds^2 = -dt^2 + a^2(t) e^{2\zeta(\hat{r})} [d\hat{r}^2 + \hat{r}^2 d\Omega^2]
\end{equation}
where $\zeta(\hat{r})$ is the comoving curvature perturbation,  $a(t)$ is the scale factor and $t$ denotes the cosmic time.  The compaction function, $\mathcal{C}(r, t)$, a robust measure of the PBH collapse probability, is defined as twice the local mass excess relative to the unperturbed background, divided by the areal radius $R(r,t)$ ~\cite{Shibata:1999zs, Musco:2018rwt},
\begin{equation}
    \mathcal{C}(r,t) \equiv 2 \frac{M(r,t) - M_b(r,t)}{R(r,t)}
\end{equation}
On superhorizon scales, where pressure gradients are negligible, the compaction function is time-independent. It can be related directly to the initial curvature perturbation $\zeta(\hat{r})$ through the gradient expansion approach, which gives~\cite{Musco:2018rwt},
\begin{equation}
    \mathcal{C}(\hat{r}) = -\Phi_{\text{EoS}}\hat{r}\zeta'(\hat{r})[2+\hat{r}\zeta'(\hat{r})] \, .
    \label{zeta_C}
\end{equation}
Where the prime denotes a derivative with respect to the comoving radius $\hat{r}$ and the function $\Phi_{\text{EoS}}(t)$ is determined by the equation of state of the Universe $w$. For Hubble radius $R_H$, we can write~\cite{Polnarev:2006aa}, 
\begin{equation}
    R_H\frac{\d\Phi_{\text{EoS}}}{\d R_H} + \frac{5+3 w}{3(1+w)} \Phi_{\text{EoS}} - 1 = 0 \, .
    \label{eq:Phi}
\end{equation}
Assuming that in past infinity  $t, R_H \to 0$, we can solve for $\Phi_{\text{EoS}}(t)$. For a constant equation of state $w$, we get,
\begin{equation}\label{solPhi}
\Phi_{\text{EoS}} = \frac{3(1+ \bar w)}{(5+3 \bar w)} \, .
\end{equation}
For radiation domination $ w = 1/3$, which sets $ \Phi=2/3$ . The smoothed overdensity $\delta(r,t)$ is estimated as the excess of mass averaged over a spherical volume of radius $R(r,t)$,  at the time of the horizon crossing,
\begin{equation}
    \delta(r,t) \equiv \frac{4\pi}{V} \int_0^R 
    \frac{\delta\rho}{\rho_b} R^2 \d R \,.
    \label{eqn:density}
\end{equation}
At the maxima of the
compaction function (\mbox{$\mathcal{C}'(r_m) = 0$}),
\bea
\zeta'(r_m) +\zeta''(r_m)=0 \, ,
\eea
and we can relate  $\mathcal{C}(r)$  to the energy density profile \cite{Musco:2018rwt}: 
\begin{equation}
    \mathcal{C}(r) = \frac{\tilde{r}^2}{\tilde{r}^2_\mathrm{m}} \delta(r,t_H) \, .
\end{equation}
Thus, at $\tilde{r}=\tilde{r}_\mathrm{m}$ the amplitude of $\delta$ can be defined as the peak value of the compaction function, $\delta \equiv \mathcal{C}_{\text{max}}$. This peak value is equivalent to the mass excess averaged over the volume of the perturbation at the time of horizon crossing, justifying its use as the criterion for collapse.
The critical threshold of $\delta$ for the formation of PBHs is defined as $\delta_c$. In RD, we can use a numerical fit from simulations to compute $\delta_{c, rad}$ ~\cite{Escriva:2019phb, Musco:2020jjb}:
\be \label{eq:delta_c_analytical_Musco}
\delta_{\mathrm{c}, rad}= 
\begin{cases}
\alpha^{0.047}-0.50 \quad\quad\quad 0.1 \lesssim\alpha\lesssim 7 \\
\alpha^{0.035}-0.475 \quad\quad\quad 7 \lesssim\alpha\lesssim 13 \\
\alpha^{0.026}-0.45 \quad\quad\quad 13 \lesssim\alpha\lesssim 30
\end{cases}
\ee
where the shape parameter $\alpha$, is defined by
\begin{equation}
    \alpha \equiv -\frac{\mathcal{C}''(\tilde{r}_m)\tilde{r}_m^2}{4\mathcal{C}(\tilde{r}_m)}
\end{equation}

\noindent Assuming curvature perturbation to be Gaussian, we can relate the shape parameter $\alpha$ to the properties of the primordial power spectrum. As we limit ourselves to a scale invariant scalar power spectrum peak, with a spectral index $n_0 = 1$, we use the shape parameter $\alpha\simeq3$, having a value of the threshold $\delta_{c, rad}\simeq0.55$ \cite{Franciolini:2022tfm}. 

From Eqn.  \eqref{zeta_C}, it is evident that $\delta$ has a non-linear dependence on the scalar curvature perturbation; thus, even if we start with a purely Gaussian curvature perturbation defined solely by its power spectrum, the density contrast $\delta$ shall have a non-Gaussian component. It is often customary to obtain a threshold for the Gaussian or linear component of $\delta$, namely $\delta_l$, to simplify the PBH abundance estimation. We can relate the linear perturbation $\delta_l \equiv -2\Phi_{\text{EoS}}\, r\zeta^\prime(r_m)$ to the physical overdensity $\delta$ as,
\be %\label{delta_l}
\delta \approx
\delta_l - \frac{1}{4 \Phi_{\text{EoS}}} \delta_l^2 . 
\label{lin2NL}
\ee

\paragraph{The mass of the PBHs and the Linear Threshold}
The formation of a PBH is a manifestation of critical collapse. For a given perturbation shape, there exists a critical threshold for the non-linear density contrast, $\delta_c$. Perturbations with amplitudes $\delta > \delta_c$ will collapse to form a black hole, while those with $\delta < \delta_c$ will disperse. For perturbations marginally above the threshold, the final PBH mass follows a scaling law ~\cite{PhysRevLett.70.9, Niemeyer:1997mt, Niemeyer:1999ak, Musco:2012au},
\begin{equation}
    M_{\text{PBH}} = K M_H (\delta - \delta_c)^\gamma \, .
    \label{eq:critical_collapse_full}
\end{equation}
Here, $M_H$ is the horizon mass at the time of collapse, the exponent $\gamma$ depends on the equation of state ($w$), and the scaling parameter $K$ depends on both $w$ and the shape of the initial perturbation profile. Though $\delta$ determines the collapse, from the inflationary scalar power spectrum, we estimate the linear variable  $\delta_l$. By solving Eq. \ref{eq:critical_collapse_full} for $\delta$ and substituting it into Eq. \eqref{lin2NL}, we can solve for the required linear threshold $\delta_l$ needed to form a PBH of mass $M_{\text{PBH}}$ from a horizon of mass $M_H$.
\begin{equation}
    \delta_l(M_H, M_{\text{PBH}}) = 2 \Phi_{\text{EoS}}(M_H) \left( 1 - \sqrt{1 - \frac{\delta_c(M_H)}{\Phi_{\text{EoS}}(M_H)} - \frac{1}{\Phi_{\text{EoS}}(M_H)}\left(\frac{M_{\text{PBH}}}{K M_H}\right)^{1/\gamma}} \right)
\end{equation}
Defining the term under the square root as $\lambda$, we need $\lambda \ge 0$, as the physical condition for PBH formation. We take $\delta_c$,  $\gamma$ and $\mathcal{K}$ as a function of $M_{\rm H}$ during the QCD phase transition as estimated in ~\cite{Musco:2023dak}.

\paragraph{Peaks theory }
The number density of peaks in the linear density field, within a range $\mathrm{d}\delta_l$ around a value $\delta_l$, is given by the Bardeen-Bond-Kaiser-Szalay (BBKS) formula \cite{Bardeen:1985tr}:
\begin{equation}
    n(\delta_l, R) = \frac{1}{(2\pi)^2} \left( \frac{\sigma_1^2(R)}{3\sigma_2^2(R)} \right)^{3/2} \frac{\delta_l^3}{\sigma_0^3(R)} \exp\left(-\frac{\delta_l^2}{2\sigma_0^2(R)}\right)
\end{equation}
where $\sigma_j^2$ are the  $j^\mathrm{th}$-order moments of the power spectrum,
\begin{equation}
    \sigma_j^2(r) = \frac{4}{9}\Phi_{\text{EoS}}^2 \int\limits_0^\infty \frac{\mathrm{d}k}{k}\left(k R \right)^4 k^{2j} \tilde{W}^2(k,R) \mathcal{P}_\zeta(k) \, .
\end{equation}
Here, $\tilde{W}(k, R)$ is the window function to coarse-grain the overdensity over a smoothing scale, which we take to be equal to the horizon size, and this corresponds to the horizon mass $M_H$. The final mass function, $f_{\text{PBH}}(M_{\text{PBH}}) = \mathrm{d}f_{\text{PBH, tot}}/\mathrm{d}\ln M_{\text{PBH}}$, is obtained by integrating over all possible horizon mass scales ($M_H$) that could form a PBH of mass $M_{\text{PBH}}$:
\begin{equation}
\begin{split}
    f_{\text{PBH}}(M_{\text{PBH}}) = M_{\text{PBH}} \int \frac{\mathrm{d}M_H}{M_H} & \left(\frac{M_{eq}}{M_H}\right)^{1/2} n(\delta_l, M_H)
     \frac{4\pi R^3(M_H)}{3} \frac{M_{\text{PBH}}}{M_H} \left(\frac{\mathrm{d}\delta_l}{\mathrm{d}M_{\text{PBH}}} \right)
\end{split}
\label{eq:fpbh_peaks_full_app}
\end{equation}
Here, $(M_{eq}/M_H)^{1/2}$ takes into account redshifting of energy density of the PBHs and the energy density of the background radiation fluid until matter-radiation equality.

%\paragraph{The Threshold Statistics}

\begin{figure}[h!]
\begin{center}
\includegraphics[scale=0.5]{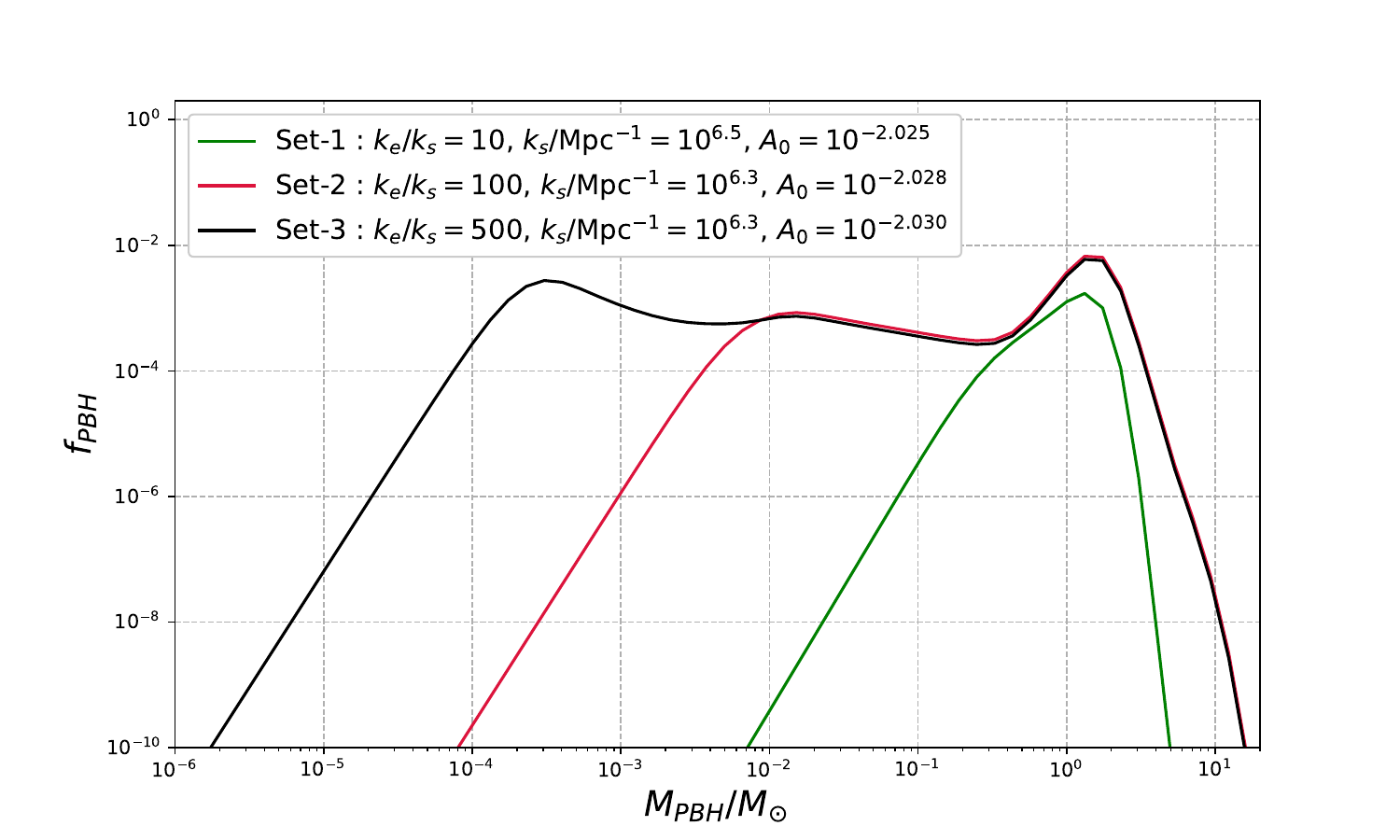}
\caption{Here we plot the PBH mass fraction $f_\text{PBH}$ in Peaks Theory (solid lines) 
%and Threshold Statistics (dashed lines) 
formalism for three different sets of parameters. These parameter choices illustrate the variation in mass fraction resulting from changes in the height, width, and location of the inflationary scalar power spectrum peak. Evidently, a broader inflationary spectrum leads to a more extended PBH mass distribution, while the change in the height of the inflationary power spectrum peak significantly changes the PBH abundance. Please note that we are limiting ourselves to exactly flat scalar power spectrum amplification $n_0=1$.}
\label{mass}
\end{center}
\end{figure}

\begin{figure}[h!]
\centering
\includegraphics[width=0.8\textwidth]{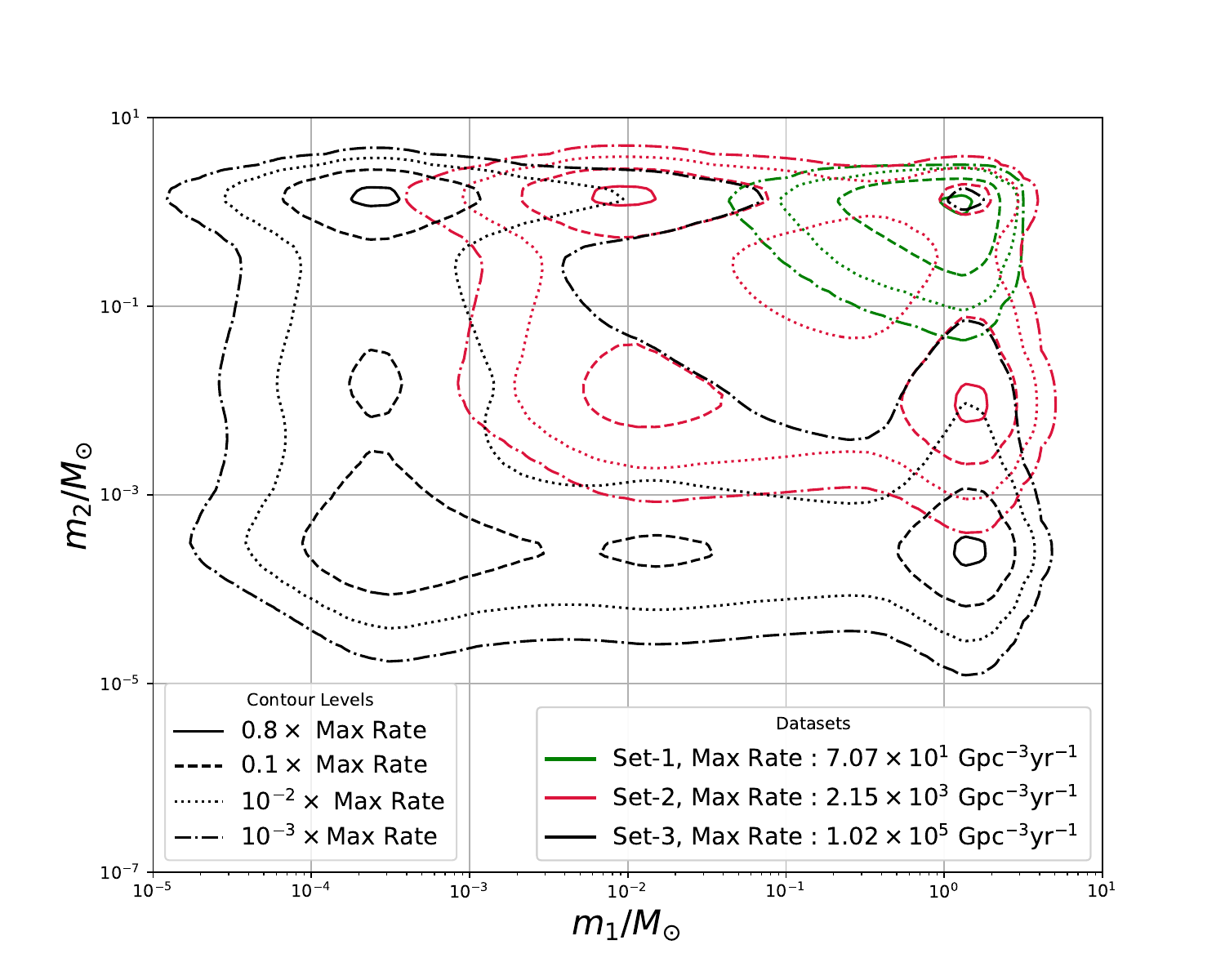}
\caption{
Here, we plot the PBH Merger Rate Density per Logarithmic Mass Interval for Different Power Spectrum Models.
The plot shows the calculated merger rate density in logarithmic mass interval ${dR_{E2}}/({d(\ln m_1) d(\ln m_2)})$ [$\mathrm{Gpc}^{-3}\,\mathrm{yr}^{-1}$], as a function of the two merging black hole masses, $m_1$ and $m_2$, in solar mass units ($M_{\odot}$).  Each set of colored contours corresponds to a different set of parameters for the primordial power spectrum, as detailed in the ``Datasets" legend. We plot the contours for 4 different levels of merger rate with respect to the maximum merger rate, as described in the ``contour levels" legend.}
\label{fig:merger_rate_contours}
\end{figure}
\subsection{PBH Binary Formation and Merger Rate}
\label{PBH-merger}
To predict the gravitational-wave signal from a population of PBHs, it is essential to calculate their merger rate density. Although we work with an extended mass distribution of PBHs, we use a theoretical framework that is most accurate for a narrow mass distribution of PBHs \cite{Raidal:2024bmm, Hutsi:2020sol, Raidal:2018bbj, Liu:2018ess, Vaskonen:2019jpv}. Several channels contribute to the total PBH merger rate, including early- and late-Universe formation mechanisms involving two- or three-body dynamics \cite{Raidal:2024bmm}. In this work, we focus on the \textbf{early two-body channel}, which is the dominant source of mergers in scenarios where PBHs constitute a subdominant fraction of dark matter ($f_{\text{PBH}} \lesssim 0.1$), the parameter space of our primary interest.

Binaries in this channel originate from pairs of PBHs in the early Universe that decouple from cosmic expansion due to their self-gravity. The weak tidal torques from surrounding inhomogeneities grant them small initial angular momenta, resulting in highly eccentric orbits ($e \approx 1$). The time until merger, or coalescence time $\tau$, is acutely sensitive to this angular momentum, as shown by Peters (1964) \cite{Peters:1964zz}:
\begin{equation}
    \tau = \frac{3}{85} \frac{r_a^4 j^7}{\eta M^3}.
\label{eq:coalescence_time}
\end{equation}
Here, $r_a$ is the semi-major axis, $j$ is the dimensionless angular momentum, $M = m_1+m_2$, and $\eta = m_1m_2/M^2$. A correction to this formula gives \cite{Raidal:2024bmm},
\be
    \tau 
    = \frac{3 r_a^4 j^7}{85 \eta M^3}
    \frac{F(j)-F(1) j}{ \left(1-\frac{121}{425}j^2\right)^{3480/2299} \left(1-j^2\right)^{24/19}}
    \approx \frac{3 r_a^4 j^7}{85 \eta M^3} \frac{1 - 0.70 j + 0.62 j^2}{1 + 0.67 j}\,,
\ee
where $F(j) \equiv F_1\left(-1/2;-1181/2299,-5/19;1/2;j^2 121/425,j^2\right)$ with $F_1$ the first Appell hypergeometric function.
\begin{figure}[htbp!]
    \centering
    
    % --- Row 1 ---
    \begin{subfigure}{0.49\textwidth}
        \includegraphics[width=\linewidth]{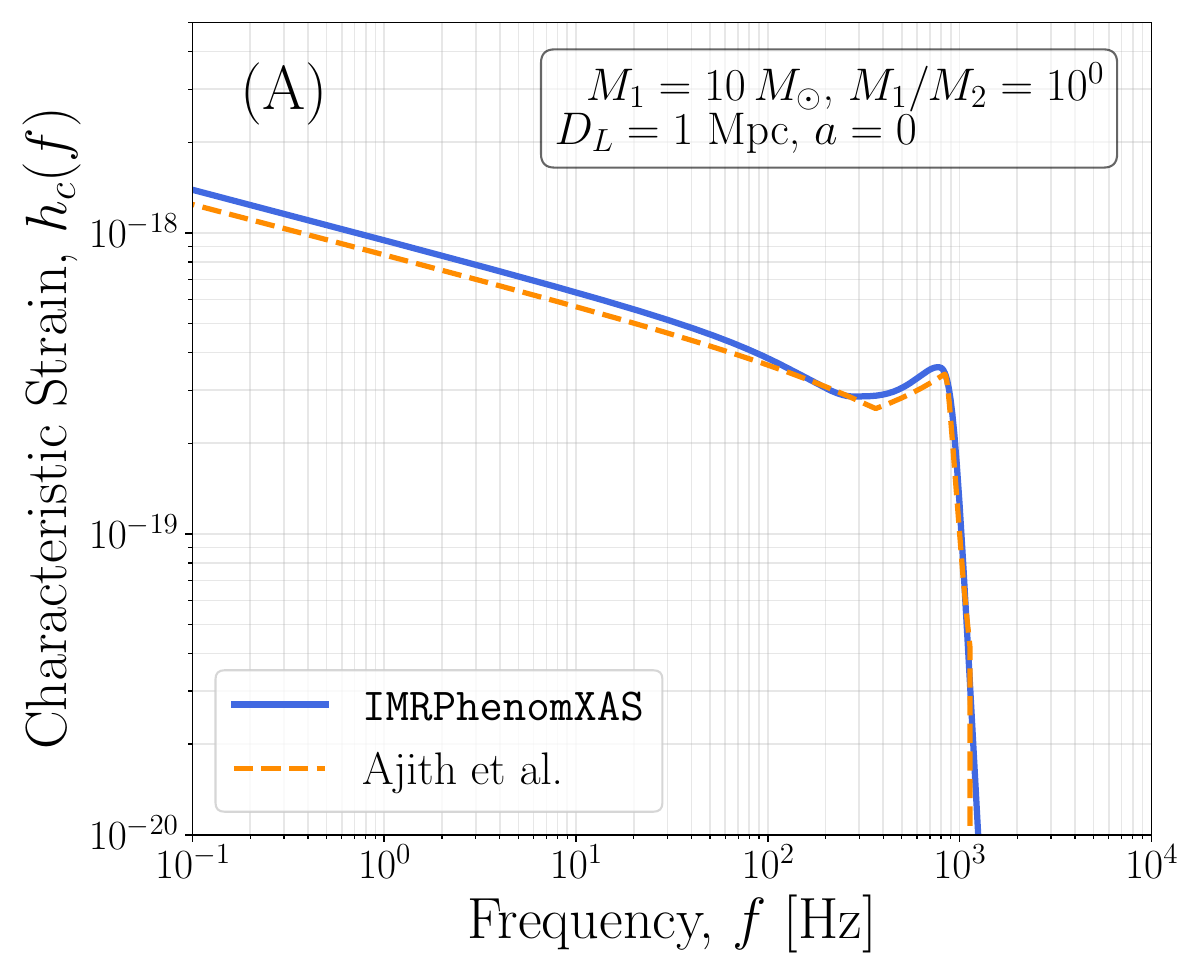}
        \label{fig:q1e0}
    \end{subfigure}
    \hfill 
    \begin{subfigure}{0.49\textwidth}
        \includegraphics[width=\linewidth]{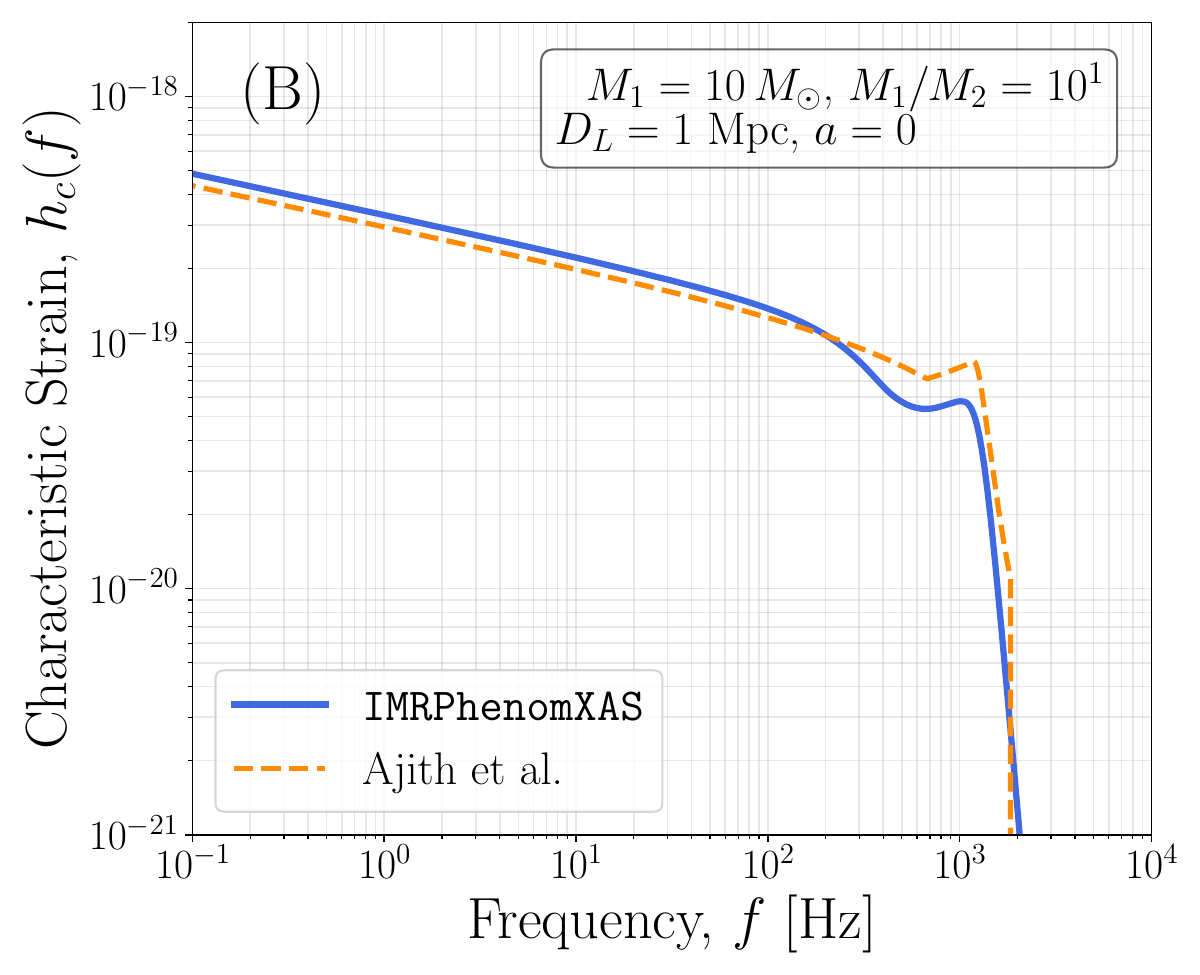}
        \label{fig:q1e1}
    \end{subfigure}

    \vspace{0.3cm} 

    % --- Row 2 ---
    \begin{subfigure}{0.49\textwidth}
        \includegraphics[width=\linewidth]{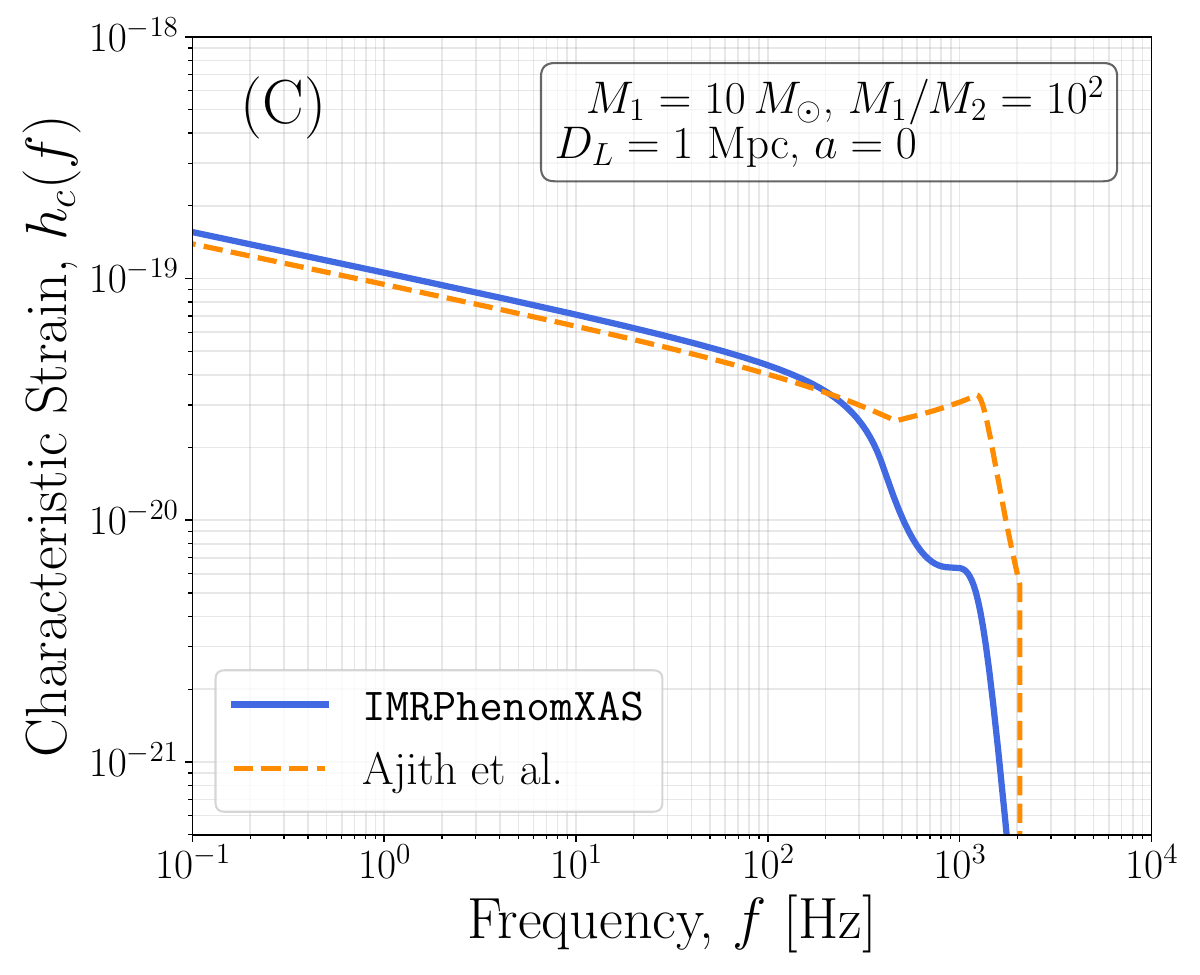}
        \label{fig:q1e2}
    \end{subfigure}
    \hfill
    \begin{subfigure}{0.49\textwidth}
        \includegraphics[width=\linewidth]{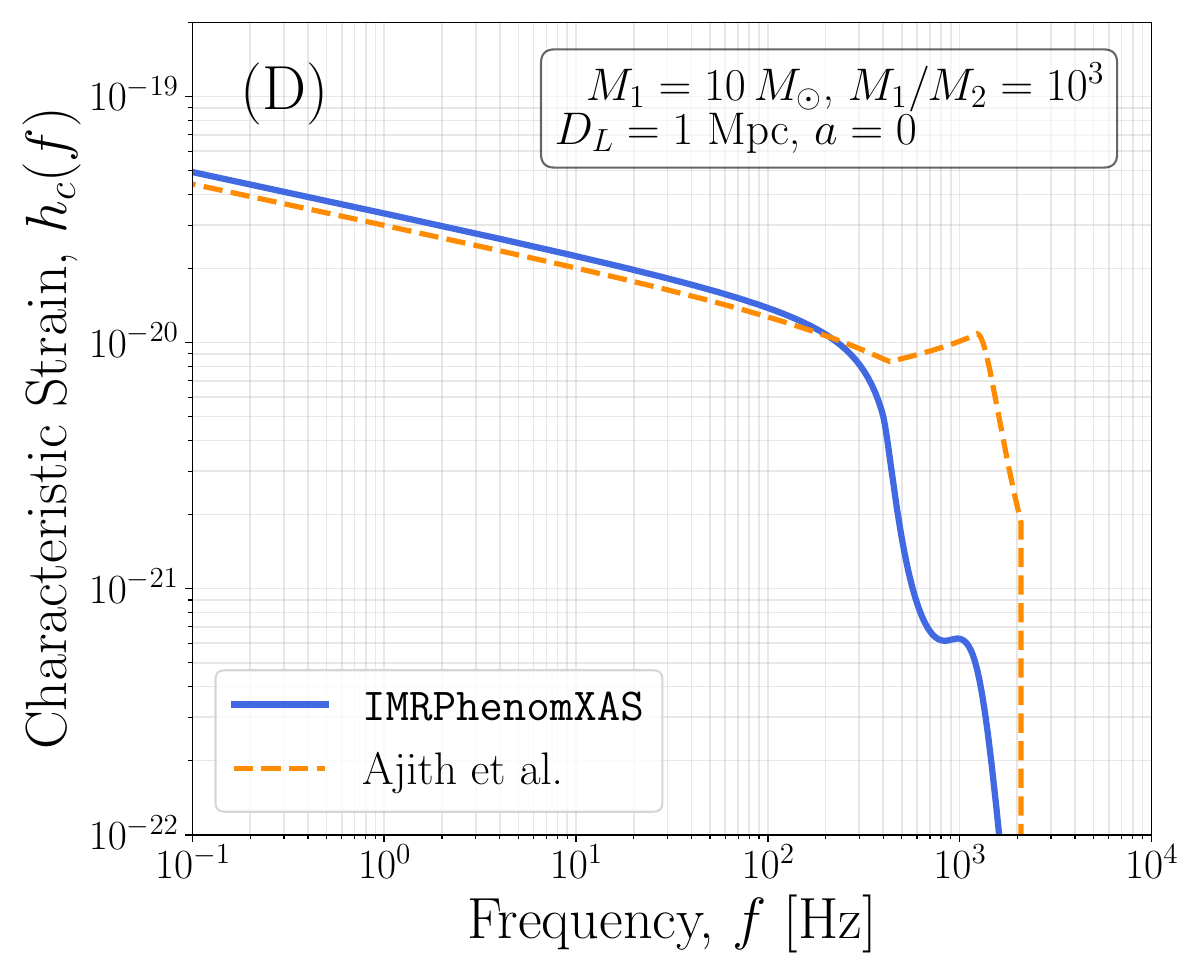}
        \label{fig:q1e3}
    \end{subfigure}

    \vspace{0.3cm}

    % --- Row 3 ---
    \begin{subfigure}{0.49\textwidth}
        \includegraphics[width=\linewidth]{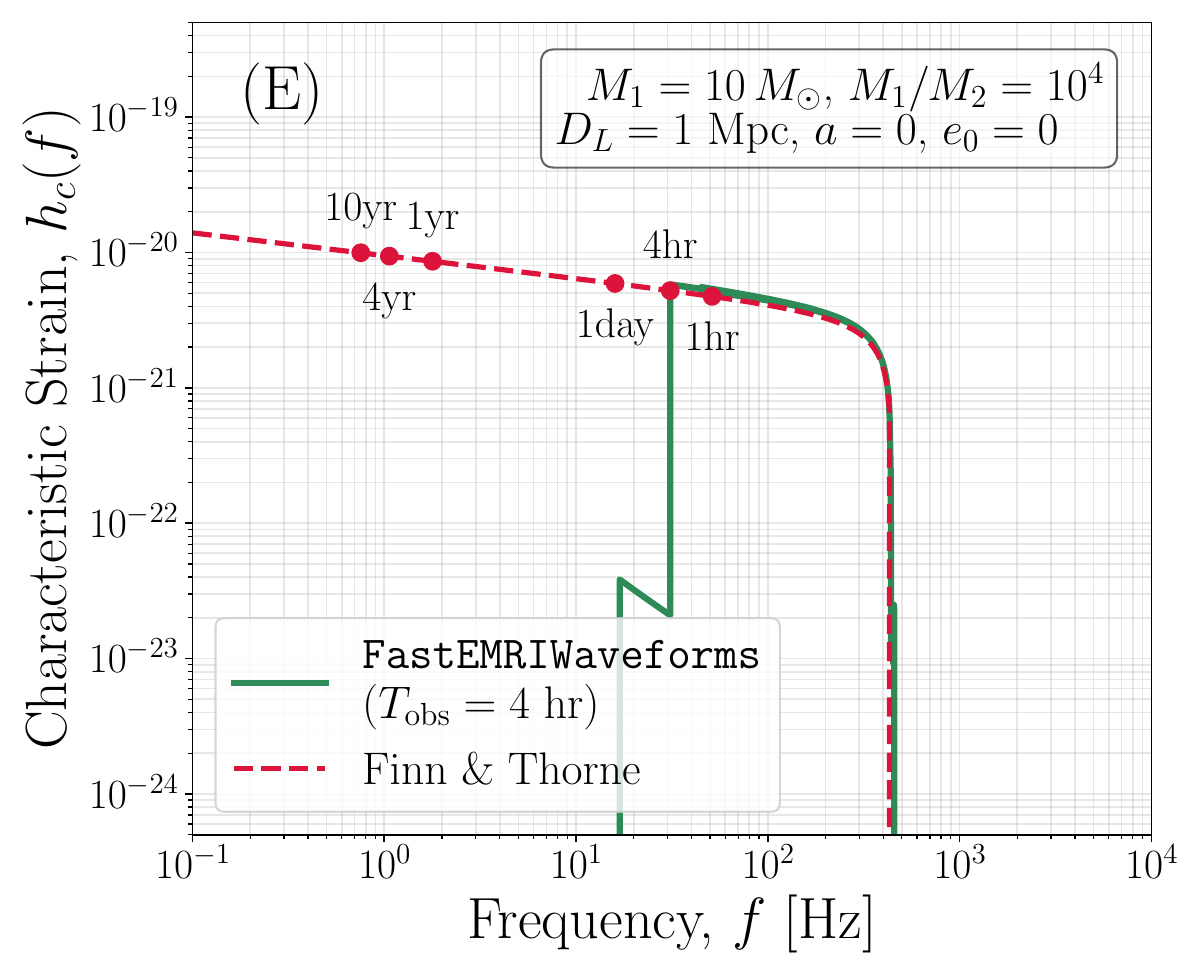}
        \label{fig:q1e4}
    \end{subfigure}
    \hfill
    \begin{subfigure}{0.49\textwidth}
        \includegraphics[width=\linewidth]{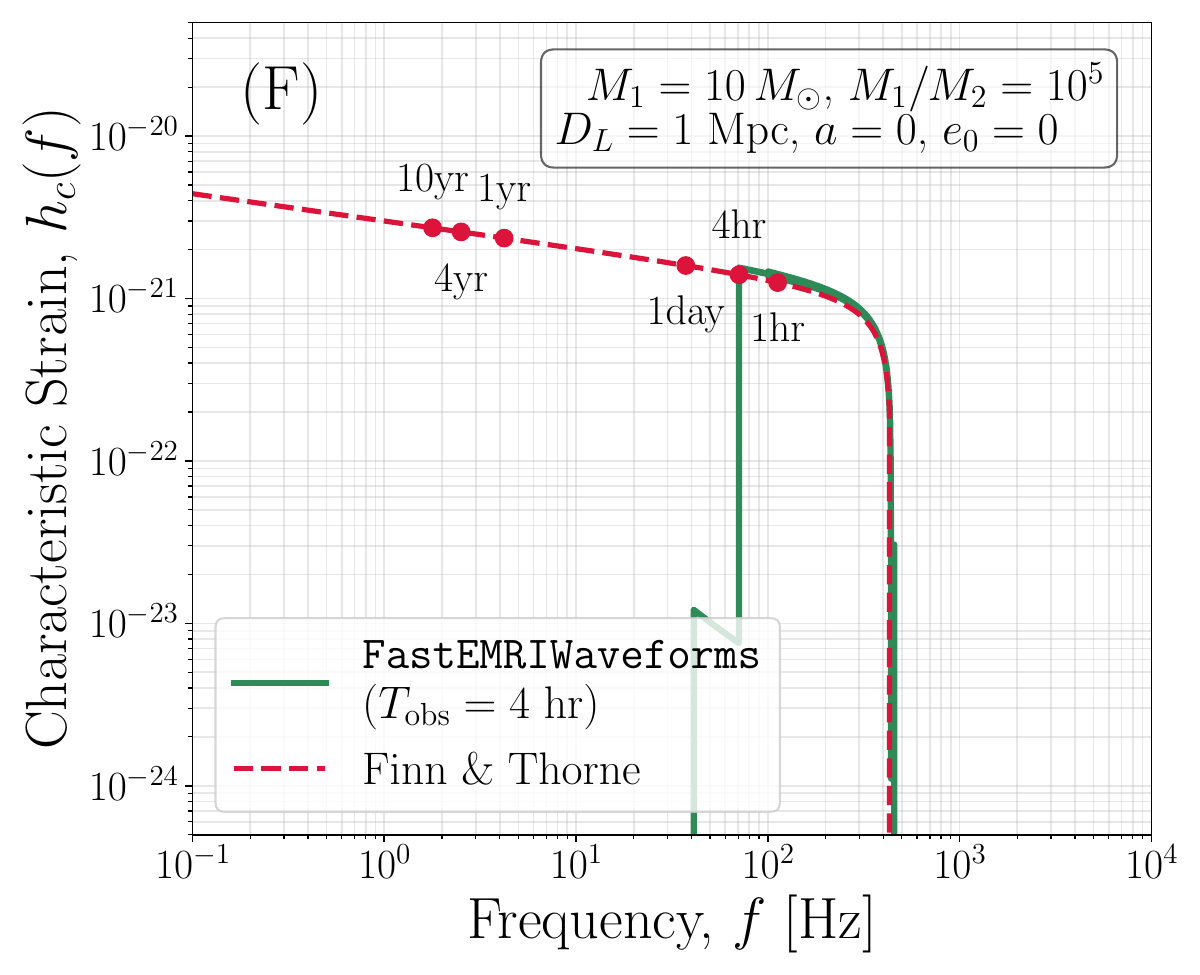}
        \label{fig:q1e5}
    \end{subfigure}

    \caption{Comparison of waveform models for a binary system with a primary mass of $M_1 = 10\, M_\odot$ at a distance of 1 Mpc. Each panel shows a different mass ratio, $Q = M_1/M_2$. Panels (A)-(D) compare the \texttt{IMRPhenomXAS}~\cite{Pratten:2020fqn} model with \cite{Ajith:2009bn}. In panels (E) and (F), we compare the mini-EMRI waveform obtained from \texttt{FastEMRIWaveforms}~\cite{Chua:2020stf, Katz:2021yft, Speri:2023jte, Chapman-Bird:2025xtd} with Eq. \eqref{eq:CharStrain} for m=2, which follows from  ~\cite{Finn:2000sy} which also takes into account the relativistic correction factors based on the Teukolsky-Sasaki-Nakamura formalism~\cite{Teukolsky:1973ha, Sasaki:1981sx}.}
    \label{fig:all_comparisons}
\end{figure}
\noindent Our calculation begins with the normalized PBH mass function, $\psi(m)$, given by:
\begin{equation}
    \psi(m) \equiv \frac{m}{\rho_{\text{PBH}}} \frac{dn_{\text{PBH}}}{d\ln m}, \quad \text{such that} \quad \int \psi(m) d\ln m = 1.
    \label{eq:mass_function}
\end{equation}
The two functions, $f_{\rm PBH}(m)$ and $\psi(m)$, are related by a simple normalization,
\begin{equation}
	\psi(m) \equiv \frac{f_{\rm PBH}(m)}{f_{\rm PBH, tot}} \, ,
\end{equation}
where the total abundance of PBHs today, $f_{\rm PBH, tot}$, is defined as:
\begin{equation}
f_{\rm PBH, tot}= \int\limits_{M_{\rm min}}^{M_{\rm max}}\mathrm{d}\ln(M)f_{\rm PBH}(M) \, .
\end{equation}
The total merger rate, $R_{E2}$, is the product of a baseline (unsuppressed) rate, $R^{(0)}_{E2}$, and two suppression factors, $S_E$ and $S_L$. We can express the baseline rate as \cite{Raidal:2024bmm, Hutsi:2020sol,Raidal:2018bbj,Vaskonen:2019jpv},
\begin{equation}
\frac{dR^{(0)}_{E2}}{d(\ln m_1) d(\ln m_2)} \approx \frac{1.6 \times 10^6}{\text{Gpc}^3 \text{yr}} f_{\text{PBH}}^{\frac{53}{37}} \eta^{-\frac{34}{37}} \left(\frac{M}{M_{\odot}}\right)^{-\frac{32}{37}}  \left(  \frac{t}{t_0} \right )^{-34/37} \psi(m_1)\psi(m_2) .
\label{eq:rate_unsuppressed_final_app}
\end{equation}
For the mergers happening at low redshifts, we can take the present-day limit $t \to t_0$.
The suppression factors account for disruptive gravitational encounters. The first factor, \textbf{$S_E$}, corrects for early-time disruption. It represents the probability that a nascent binary is sufficiently isolated from a third PBH to avoid a chaotic three-body interaction. It is approximated by \cite{Raidal:2018bbj}:
\begin{equation}
    S_E \approx \frac{\sqrt{\pi}(5/6)^{21/74}}{\Gamma(29/37)} \left[ \frac{\langle m^2 \rangle / \langle m \rangle^2}{\bar{N}(y) + C} + \frac{\sigma_M^2}{f_{\text{PBH}}^2} \right]^{-\frac{21}{74}} e^{-\bar{N}(y)},
    \label{eq:se_final}
\end{equation}
where $\sigma_M^2 \approx 0.005$ is the variance of matter density, and $\bar{N}(y)$ is the expected number of nearby disruptive PBHs:
\begin{equation}
    \bar{N}(y) \approx \frac{M}{\langle m \rangle} \frac{f_{\text{PBH}}}{f_{\text{PBH}} + \sigma_M}.
\label{eq:nbar_final}
\end{equation}
Conversely, the factor \textbf{$S_L$} accounts for late-time perturbations occurring within the dense PBH halos that form as cosmic structure grows. While these encounters rarely break the hard binaries, they can increase their angular momentum, thereby pushing their merger far into the future. This suppression is modeled as \cite{Vaskonen:2019jpv}:
\begin{equation}
    S_L \approx \min\left\{1, \; 0.01 f_{\text{PBH}}^{-0.65} e^{0.03 \ln^2(f_{\text{PBH}})}\right\}.
\label{eq:sl_final}
\end{equation}
The final merger rate used in our analysis is the product of these three components:,
\begin{equation}
\frac{dR_{E2}}{d(\ln m_1) d(\ln m_2)}  = \frac{dR^{(0)}_{E2}}{d(\ln m_1) d(\ln m_2)}  \times S_E \times S_L.
\end{equation}
We shall also use the merger rate per linear mass intervals,
\begin{equation}
\frac{dR_{E2}}{d m_1 d m_2}  = \frac{dR^{(0)}_{E2}}{d(\ln m_1) d(\ln m_2)} \frac{1}{ m_1 m_2} \, .
\label{eq:merg_lin}
\end{equation}
\section{Probing the Scenario with SGWB}
\label{sec:sgwb_probes}
\begin{figure}[htbp!]%[!t]%[t!]
\begin{center}
\includegraphics[scale=0.55]{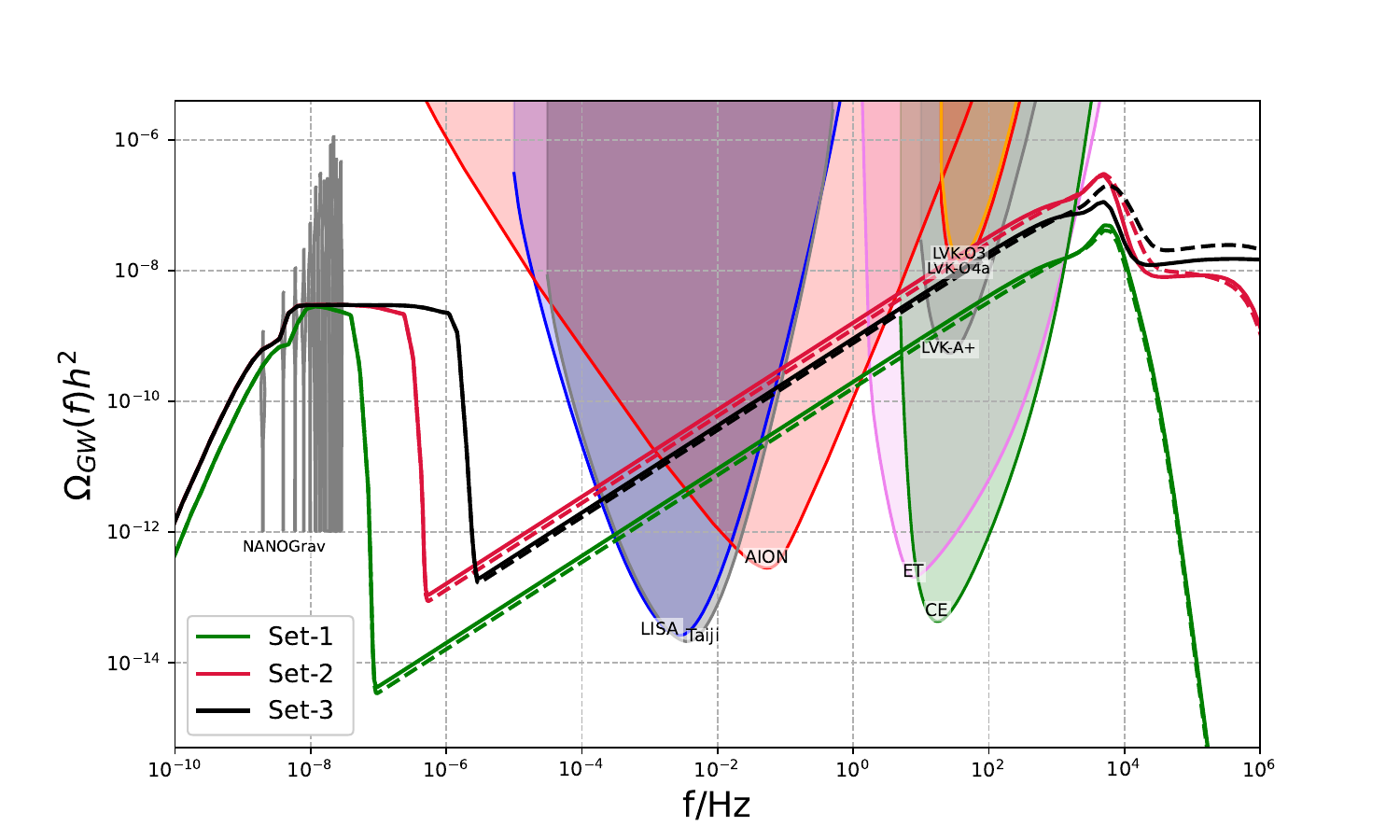}
\caption{We plot the SGWB spectra for three different sets of parameters. The amplification in SGWB in the left part of the plot is for second-order tensor perturbation during the PBH formation, while the right side of the plot corresponds to the SGWB from PBH mergers. The solid SGWB lines make use of \texttt{IMRPhenomXAS}~\cite{Pratten:2020fqn} waveform of \texttt{pycbc}~\cite{alex_nitz_2024_10473621} for  $M_1/M_2 < 10^3$ and EMRI waveform \cite{Finn:2000sy} for $M_1/M_2 > 10^3$ while the dashed lines correspond to using \cite{Ajith:2009bn} for all mass ratios. Here we also plot the Kernel Density Estimations (KDE) for NANOGrav,  $2\sigma$  power law spectrum sensitivity lines for LVK O3 and O4a~\cite{LIGOScientific:2025bgj} and projected power law integrated sensitivity curves~\cite{Schmitz:2020syl} for LVK A+, \cite{LIGOScientific:2016fpe, LIGOScientific:2019vic, KAGRA:2021kbb, LIGO-G2001287}, ET~\cite{Punturo:2010zz, Hild:2010id}, LISA~\cite{Bartolo:2016ami, Caprini:2019pxz, LISACosmologyWorkingGroup:2022jok, Schmitz:2020rag}, AION~\cite{Badurina:2019hst} CE~\cite{Evans:2021gyd} and ET~\cite{Punturo:2010zz, Hild:2010id} and Taiji~\cite{Luo:2021qji}  with a cutoff SNR $\ge$ 10 and assuming operation time $\mathcal{T}=4$~yr for LISA, Taiji, CE and ET; 1 yr for LVK detectors.}
\label{omega}
\end{center}
\end{figure}
%\FloatBarrier
Our model, characterized by an extended PBH mass function, naturally predicts two distinct and observationally separated stochastic gravitational-wave backgrounds. The first is a low-frequency background generated as a second-order tensor perturbation sourced by enhanced scalar perturbation, during PBH formation, and the second is a higher-frequency background produced by the superposition of countless PBH binary mergers throughout cosmic history. In this section, we detail the calculation of both signals and present the results of our analysis comparing them to experimental data.

\subsection{SGWB Spectra and SNR}

\subsubsection{The Second-Order Induced SGWB}
The highly amplified first-order scalar perturbations, required for PBH formation, in turn, act as a source for tensor perturbations at second order, generating a distinct SGWB \cite{Mollerach:2003nq, Ananda:2006af, Baumann:2007zm, Espinosa:2018eve, Kohri:2018awv}. The energy density of this induced background per logarithmic comoving wavenumber ($k$) interval, $\Omega_{\rm GW}(\tau,k)$, is given by:
\begin{align}
\label{omega_uv}
\Omega_{\rm GW}^{\rm form}(\tau,k)= 
\frac{1}{6}  \int_{0}^{\infty}dv \int_{|1-v|}^{1+v} du
\left(\frac{4 v^2-(1+v^2-u^2)^2}{4 u v}\right)^2 \nonumber\\ 
\times \hspace{0.5cm}\overline {\cal I}^2(v,u,x) {\cal P}_{\cal R}(kv){\mathcal P}_{\mathcal R}(ku) \hspace{1cm}\, .
\end{align}
Here, ${\mathcal P}_{\mathcal R}$ is the inflationary scalar power spectrum (Eq.~\ref{inflPS}), and the kernel term $\overline{\cal I}^2$ describes the evolution of the scalar modes after they re-enter the horizon. It is important to note that for the second-order SGWB calculation, we assume RD even during the QCD phase transition, as the softening of the equation of state would lead to negligible modifications. We calculate the spectrum long after PBH formation, during the radiation-dominated (RD) era, by which time the kernel saturates to its final form \cite{Espinosa:2018eve, Kohri:2018awv}:
\begin{align}
\overline {\cal I}_{\rm RD}^2(v,u,x \to \infty) 
=  \frac{1}{2}\l(\frac{3 (u^2+v^2-3)}{4u^3v^3}\r)^2\nonumber \\
\biggl[\l( -4uv +(u^2+v^2-3)\,{\rm log}\l|\frac{3-(u+v)^2}{3-(u-v)^2}\r|\r)^2 \biggr. \nonumber \\
 \biggl. \!\! +\pi^2 (u^2+v^2-3)^2 \Theta(u+v-\sqrt{3})\biggr].
 \label{ird}
\end{align}
The present-day SGWB spectrum, $\Omega_{\rm GW}^{\rm form}(\tau_0,k)$, is then estimated by redshifting this result from the time of formation to today:
\begin{align}
\Omega_{\rm GW}^{\rm form}(\tau_0,k) =  c_g ~\Omega_{r,0} ~\Omega_{\rm GW}^{\rm form}(\tau,k)\, ,
\label{oform}
\end{align}
where $\Omega_{r,0}$ is the present-day radiation density and $c_g$ is a factor accounting for the change in relativistic degrees of freedom.
\subsubsection{SGWB from PBH Binary Mergers}
The second contribution to the SGWB comes from the superposition of gravitational waves from the multitude of PBH-PBH mergers. The energy density spectrum is calculated by integrating the energy emitted per merger over the cosmic merger rate history \cite{Phinney:2001di, Zhu:2011bd, Rosado:2011kv, Wang:2016ana}:
\begin{equation}
\Omega_{\text{GW}}^{\rm merger}(f_{\text{fre}},m_1,m_2)=\frac{f_{\text{fre}}}{\rho_cH_0}\int_{z_{min}}^{z_{\text{max}}}\frac{dz}{(1+z)E(\Omega_{\text{M}},\Omega_\Lambda,z)} \times \frac{dR_{E2}}{d m_1 d m_2}\Big\vert_{t=t(z)} \times \frac{dE_{\text{GW}}(f_s,z)}{df}\, ,
\label{sgwbmer}
\end{equation}
and to obtain the total spectrum, we integrate over all contributing binary masses,
\begin{equation}
    \Omega_{\text{GW}}^{\rm merger}(f_{\text{fre}})=\int dm_1dm_2\Omega_{\text{GW}}(f_{\text{fre}},m_1,m_2).
\end{equation}

\noindent Here, we use the differential merger rate calculated in Eq.~\eqref{eq:merg_lin}. 
The quantity $\frac{{\rm d}E_{\text{GW}}}{{\rm d}f}$ denotes the spectral energy density emitted by a single binary~\cite{Ajith:2009bn}, 
$f_{\text{fre}}$ is the observed frequency, and $f_s \equiv f_{\text{fre}}(1+z)$ is the corresponding source-frame frequency. 
The frequency-domain waveform amplitudes at the source and at the detector are related by
$\tilde{h}_{\text{obs}}(f_{\text{fre}}) = \tilde{h}_{\text{src}}(f_s)/D_L$, 
where $D_L$ is the luminosity distance. 
The spectral energy distribution is related to the observed waveform through the general relation~\cite{Flanagan:1997sx, Phinney:2001di, Maggiore:2007ulw, Maggiore:2018sht},
\begin{equation}
    \frac{{\rm d}E_{\rm GW}}{{\rm d}f}
    = \frac{\pi c^3}{2G}\, f^2 D_L^{\,2}
      \int {\rm d}\Omega \,
      \big(|\tilde{h}_+(f)|^2 + |\tilde{h}_\times(f)|^2\big)
    \simeq \frac{16\pi^2 c^3}{5G}\, f^2 D_L^{\,2} |\tilde{h}(f)|^2,
    \label{eq:dEdF_withDL}
\end{equation}
where $\tilde{h}(f)$ refers to the observed strain amplitude, and the second expression assumes isotropic emission and averages over binary orientations. 
The corresponding \emph{characteristic strain} $h_c(f)$, which represents the strain amplitude per logarithmic frequency interval, is then defined as
\begin{equation}
    h_c(f)
    = 2 f\, |\tilde{h}(f)|
    = \frac{1}{\pi D_L}
      \sqrt{\frac{2G}{c^3}\,
      \frac{{\rm d}E_{\rm GW}}{{\rm d}f}},
    \label{eq:hc_def}
\end{equation}
thereby connecting the observable strain spectrum to the intrinsic energy radiated by a single coalescence event at luminosity distance $D_L$. 

It is very interesting to note that, though in Eq. \eqref{sgwbmer} we calculate the total SGWB contribution originating from the PBH model, taking $z_{min} = 0$, the part of this background arising from the comparatively lower redshifts would already be detectable or ruled out by the GW detector. The subtraction of the already detected background is already an active field in GW astronomy.  As shown in \cite{Regimbau:2016ike, KAGRA:2021kbb}, after removing the resolvable signals, the remaining stochastic background is expected to differ from the total background by no more than about $10\%$ for advanced LIGO and Virgo operating at design sensitivity, while for the A+ upgrade, this residual is anticipated to be roughly a factor of two lower. However, this depends on the black hole population considered. In our case, we deal with a broader mass distribution, including even very light mass PBHs. Thus, in some parts of our parameter space, we can expect even a larger fraction of the total SGWB contribution to come from the resolvable part.

There are two aspects to considering the total SGWB contribution instead of only the unresolved part. First, if the comparatively small number of sources at very low redshifts contributes to the majority of the SGWB background, the resulting signal will not be a stochastic background. Second, there is the aspect of detectability, or how the detector treats the SGWB background. In the case of LVK, the number of detected events is still not large, and the already resolved background is not subtracted in the O3 and O4a analyses~\cite{KAGRA:2021kbb, LIGOScientific:2025bgj}. However, with increased sensitivity in future ground and space-based detectors, the number of detected events will increase significantly, and it will be very important to separate the resolvable from the unresolvable parts of the SGWB. A straightforward way to estimate the unresolved part of the SGWB is to set a minimum redshift limit $z_{min}=z_{min} (m_1, m_2, f_{re})$ as a function of black hole masses and frequency, in the redshift integral of Eq. \eqref{sgwbmer}, where for $z< z_{min} (m_1, m_2, f_{re})$ the merger events will be individually detectable \cite{Sah:2025agw}. We can set an SNR cutoff to determine detectability, which we discuss further in the EMRI detection section \ref{sec:emri_probes}. This would differ for each detector, as detector sensitivity varies, and we leave further work in this direction for the future.

In Fig. \ref{fig:all_comparisons}, we show the comparison of the waveforms for different models. In case of $M_1/M_2 > 10^3$, the inspiral contributions dominate, and we adopt the waveform following from Finn $\&$ Thorne \cite{Finn:2000sy}. As shown in panels (E) and (F) of Fig. \ref{fig:all_comparisons}, this waveform matches very closely that of what we get from the numerical package \texttt{FastEMRIWaveforms}~\cite{Chua:2020stf, Katz:2021yft, Speri:2023jte, Chapman-Bird:2025xtd} assuming circular orbits.
We discuss the waveforms for extreme-mass-ratio systems in section~\ref{sec:emri_probes} in greater detail. 
As we can see from Fig.~\ref{fig:all_comparisons} and~\ref{omega}, though as we approach more asymmetric mass ratio systems, the individual waveforms differ visibly between \texttt{IMRPhenomXAS} and \cite{Ajith:2009bn}, when we compare the combined contribution in the form of SGWB energy density spectrum, the results assuming the phenomenological form \cite{Ajith:2009bn} for all mass ratios (the dashed lines), and the more careful implementation of \texttt{IMRPhenomXAS} output for  $M_1/M_2 < 10^3$ and EMRI waveform \cite{Finn:2000sy} for $M_1/M_2 > 10^3$ (the solid lines) have negligible difference. 
\subsubsection{Estimating the SNR for SGWB}
We use the combined SGWB spectra $\Omega_{\rm GW} = \Omega_{\text{GW}}^{\rm merger} + \Omega_{\rm GW}^{\rm form}$ to determine the signal-to-noise ratio (SNR) for different future GW detectors~\cite{Moore:2014lga},
\be
{\rm SNR} \equiv \sqrt{\mathcal{T}\int {\rm d}f\, \left[\frac{\Omega_{\rm GW}(f)}{\Omega_{\rm noise}(f)}\right]^2} \, .
\ee
For later analysis, we will set a cutoff SNR of 10 and assume an operational time of $\mathcal{T}=4$ yr for LISA, Taiji, CE, and ET, and 1 yr for LVK detectors.
\subsection{Results of the SGWB Analysis}

The doubly-peaked SGWB signals predicted by our model are shown in Figure~\ref{omega} for three benchmark parameter sets. The low-frequency peak from second-order perturbations falls squarely within the nHz band probed by PTAs. At the same time, the broad, higher-frequency background from mergers is a target for both ground-based detectors, such as LVK, ET, and CE, and space-based detectors, including LISA, AION, and Taiji.

\subsubsection{Bayesian Analysis of NANOGrav 15-Year data and LVK O1-O3 data}
To constrain the primordial parameter space, we performed a Bayesian analysis comparing our predicted second-order SGWB to the NANOGrav 15-year dataset and  LVK O1-O3 data. 
\begin{figure}[t!]
\begin{center}
\includegraphics[scale=0.9]{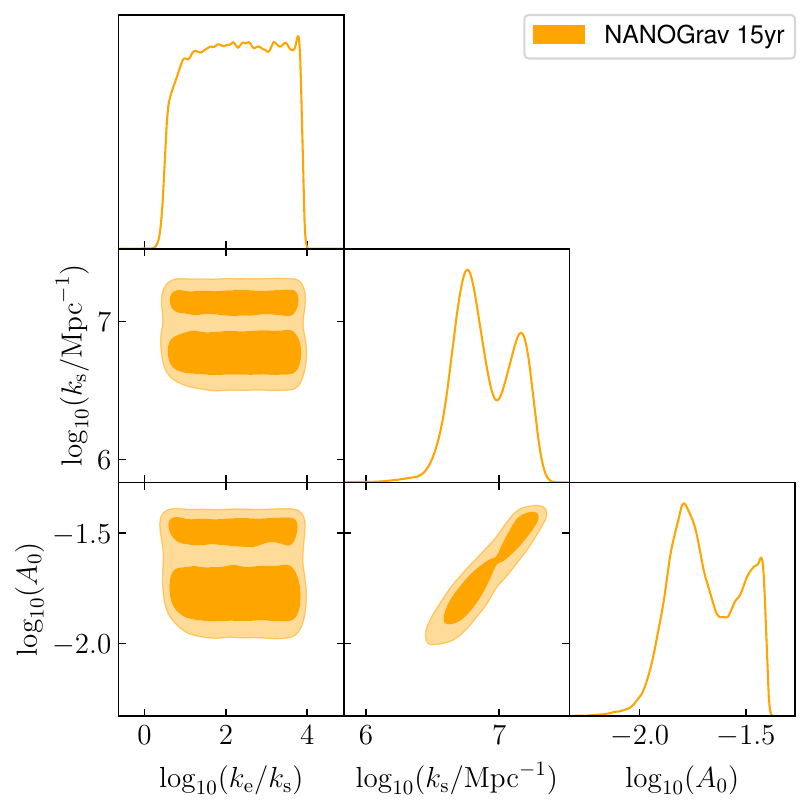}
\caption{Posterior probability distributions for the primordial power spectrum parameters: the amplitude $\log_{10}(A_0)$, the short wavenumber $\log_{10}(k_s / \mathrm{Mpc}^{-1})$, and $\log_{10} (k_e /k_s)$ for the NANOGrav 15 year data. The MCMC scan was performed with \texttt{PTArcade}~\cite{Mitridate:2023oar}, and plotted with \texttt{GetDist}~\cite{Lewis:2019xzd}.}
\label{nanograv}
\end{center}
\end{figure}
\begin{figure}[t!]
    \centering
    \includegraphics[width=0.9\linewidth]{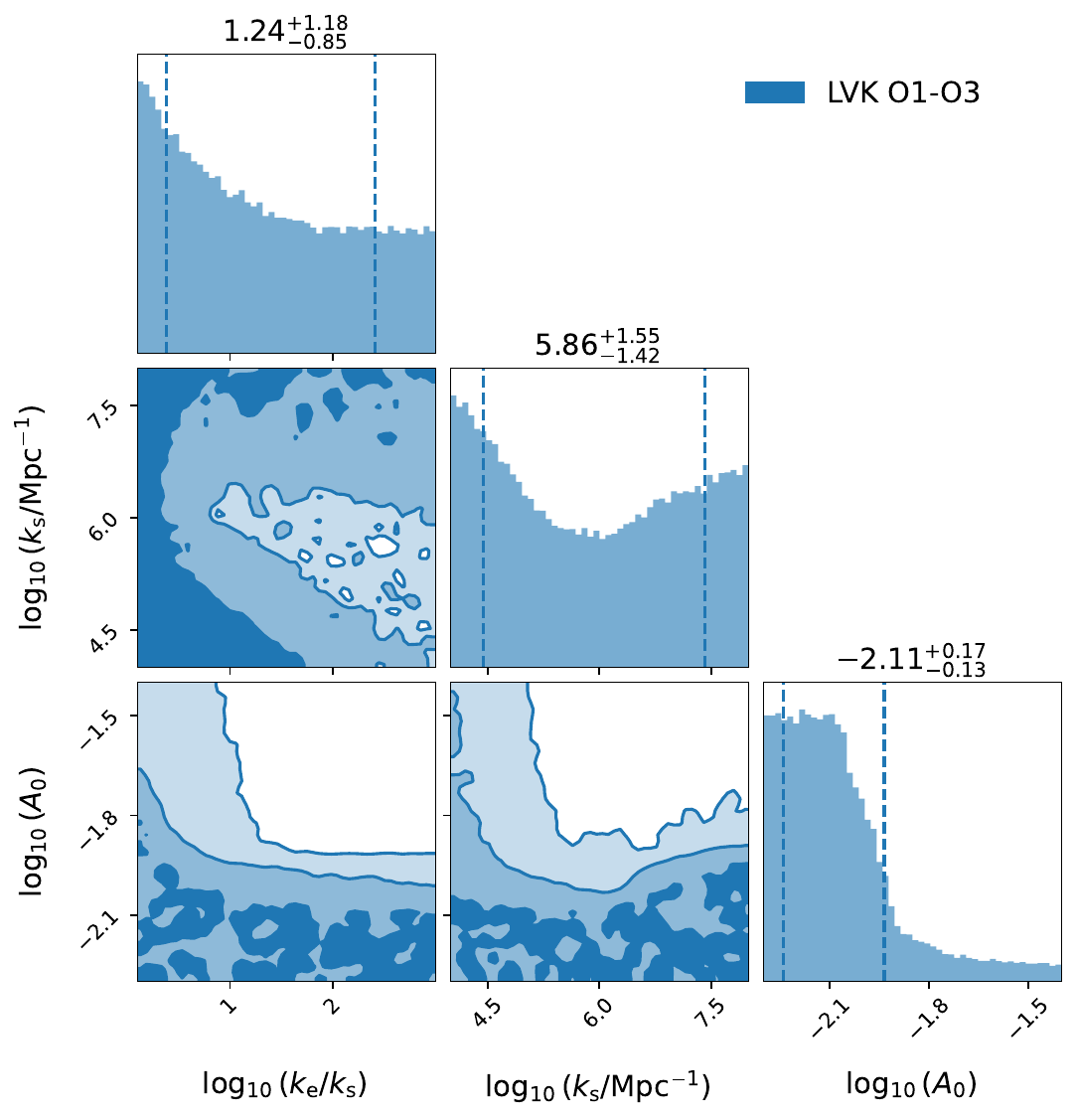}
    \caption{Posterior probability distributions for the primordial power spectrum parameters: the amplitude $\log_{10}(A_0)$ and the short wavenumber $\log_{10}(k_s / \mathrm{Mpc}^{-1})$, and the broadness parameter $\log_{10}(k_e /k_s)$. The constraints are obtained by fitting our combined SGWB spectra to a combination of data from the LVK O1, O2, and O3 observing runs, utilizing the public data \cite{KAGRA:2021kbb, LIGO-G2001287}. The contours represent the 68\% (1$\sigma$), 95\% (2$\sigma$), and 99\% (3$\sigma$) credible regions. The parameter estimation was performed with \texttt{pygwb} \cite{Renzini:2023qtj} using the \texttt{Bilby} \cite{Ashton:2018jfp} library with the \texttt{dynesty}~\cite{Speagle_2020} nested sampler.}
    \label{fig:pygwb_corner}
\end{figure}
\renewcommand{\arraystretch}{1.2} 
\begin{table}[h!]
\centering
\scriptsize
\begin{tabular}{|c|c|c|c|c|c|}
\hline
\textbf{Parameter} & \textbf{Prior} & \textbf{Mean value} & \textbf{68\% C.I.} & \textbf{95\% C.I.} & \textbf{99\% C.I.} \\
\hline
$\log_{10}\!\left(\tfrac{k_e}{k_s}\right)$ & $[0.2, 3.9]$
& $2.23$ & [1.471, 3.816] & [0.649, 3.852] & [0.477, 3.900] \\
\hline
$\log_{10}\!\left(\tfrac{k_s}{\mathrm{Mpc}^{-1}}\right)$ & $[5.6, 7.4]$
& $6.89$ & [6.632, 7.230] & [6.541, 7.296] & [6.372, 7.360] \\
\hline
$\log_{10}(A_{0})$ & $[-2.4, -1.4]$
& $-1.69$ & [-1.881, -1.412] & [-1.936, -1.401] & [-2.038, -1.389] \\
\hline
\end{tabular}
\caption{We list the priors, mean values, and the 68\%, 95\%, and 99\% confidence intervals (C.I.) for the model parameters obtained upon Bayesian analysis of the NANOGrav 15-year data with $10^7$ sampling points. The MCMC scan was performed with \texttt{PTArcade}~\cite{Mitridate:2023oar} in \texttt{ENTERPRISE}~\cite{2019ascl.soft12015E} mode, and the results were processed with \texttt{GetDist}~\cite{Lewis:2019xzd}. 
}
\label{pta_params_3p_full}
\end{table}

\begin{table}[ht!]
\centering
\scriptsize
\renewcommand{\arraystretch}{1.8}

% Adjust these widths to tune layout
\newcolumntype{L}{>{\RaggedRight\arraybackslash}p{4.5cm}} % Left model width
\newcolumntype{R}{>{\RaggedRight\arraybackslash}p{3.9cm}} % Right model width

\begin{tabular}{|c|L|c|R|c|}
\hline
\textbf{Parameter}
    & \makecell[c]{\textbf{SMBHB: 2D Gaussian prior}\\[2pt]
      \texttt{mod\_sel=True, bhb\_th\_prior=True}}
    & $\log_{10}\mathcal{B}$
    & \makecell[c]{\textbf{SMBHB: Uniform prior}\\[2pt]
      \texttt{mod\_sel=True, smbhb=True}}
    & $\log_{10}\mathcal{B}$ \\
\hline
$\log_{10}(A_{\rm SMBH})$
    & Gaussian $(-15.6,\ 0.28)$
    & \multirow{2}{*}{\textbf{$2.79 \pm 1.95$}}
    & Uniform $[-18,-14]$
    & \multirow{2}{*}{\textbf{$2.53 \pm 1.65$}} \\
\cline{1-2}\cline{4-4}
$\gamma_{_{\rm SMBH}}$
    & Gaussian $(4.7,\ 0.12)$
    &
    & $13/3$
    & \\
\hline
\end{tabular}
\caption{Comparison of the SMBHB model priors and corresponding Bayes factors obtained with \texttt{PTArcade}~\cite{Mitridate:2023oar} in \texttt{ENTERPRISE}~\cite{2019ascl.soft12015E} mode.}
\label{tab:SMBHB_prior}
\end{table}

%--- LaTeX Table Code ---
\begin{table}[h!]
\centering
\scriptsize
\begin{tabular}{|c|c|c|c|c|c|}
\hline
\textbf{Parameter} & \textbf{Prior} & \textbf{Mean value} & \textbf{68\% C.I.} & \textbf{95\% C.I.} & \textbf{99\% C.I.} \\
\hline
$\log_{10}(k_\mathrm{e}/k_\mathrm{s})$ & [0.1, 3.0] & 1.35 & [0.385, 2.411] & [0.141, 2.909] & [0.108, 2.982] \\ \hline
$\log_{10}(k_\mathrm{s}/\mathrm{Mpc}^{-1})$ & [4.0, 8.0] & 5.91 & [4.445, 7.413] & [4.065, 7.913] & [4.013, 7.983]  \\ \hline
$\log_{10}(A_0)$ & [-2.3, -1.4] & -2.07 & [-2.239, -1.935] & [-2.290, -1.542] & [-2.298, -1.432] \\
\hline
\end{tabular}
\caption{Posterior distribution values for the model parameters obtained from the analysis of LVK O1-O3 data. We list the prior ranges, mean values, and the 68\%, 95\%, and 99\% credible intervals (C.I.).
The log Bayes factor of the signal model compared to the noise-only model is $\log_{10}(\mathcal{B}) = -0.37 \pm 0.001$, indicating the data shows no preference for the signal model over Gaussian noise.}
\label{tab:lvk_param_summary}
\end{table}
For the NANOGrav 15-year data, resulting posterior distributions for the power spectrum parameters, $A_0$, $k_s$, and $k_e/k_s$, are presented in the posterior distribution plot of Figure~\ref{nanograv}. The priors and mean values and the 68\%, 95\%, and 99\% confidence intervals for the model parameters derived from this analysis are in Table~\ref{pta_params_3p_full}. The Markov Chain Monte Carlo (MCMC) scan was performed with \texttt{PTArcade}~\cite{Mitridate:2023oar} in \texttt{ENTERPRISE}~\cite{2019ascl.soft12015E} mode, and the results were processed with \texttt{GetDist}~\cite{Lewis:2019xzd}. Using the default options of \texttt{PTArcade} we modelled the expected signal produced by SMBHBs as~\cite{Phinney:2001di, Mitridate:2023oar, NANOGrav:2023hvm},
\begin{align}
    h^2\Omega_{\scriptscriptstyle\textrm{GW}}(f) = \frac{2 \pi^2A_{\scriptscriptstyle\textrm{BHB}}^2}{3 H_0^2} \left(\frac{f}{\textrm{year}^{-1}}\right)^{5-\gamma_{\scriptscriptstyle\textrm{BHB}}}\textrm{year}^{-2}\,.
\end{align}
As shown in Table \ref{tab:SMBHB_prior}, we used two distinct prior classes for the SMBHB model. The first class of priors follows a theoretical 2D Gaussian distribution of SMBHB parameters $A_{\rm SMBH}$ and  $\gamma_{_{\rm SMBH}}$~\cite{NANOGrav:2023hvm}. In the other case, we fix the tilt of the spectrum $\gamma_{_{\rm SMBH}} = 13/3$~\cite{Phinney:2001di}, and assume a log-uniform distribution for the amplitude, $A_{\rm SMBH}$. In both these cases, our Bayesian analysis favours the second-order SGWB from the PBH model over that from SMBHB mergers, indicating a strong preference for the PBH model. 
 It is essential to note that the value of the Bayes factor that we obtain with \texttt{PTArcade} has a strong dependence on both astrophysical SMBHB priors and the priors taken for PBH forming models. However, different prior choices of SMBHB do not affect the posterior distribution of PBH parameters if we keep the PBH model priors fixed. Thus, in Table~\ref{pta_params_3p_full} and Figure~\ref{nanograv}, we only show the PBH priors and posteriors obtained with 2D Gaussian priors for the SMBHB model.

As shown in Figure~\ref{nanograv}, for both $A_0$ and $k_s$, we obtain a bimodal probability distribution with two 1$\sigma$ regions. While the region with higher values of $A_0$ and $k_s$ will be subject to constraints from PBH overproduction, the second region with lower values of $A_0$ and $k_s$ is physically more interesting. 
The analysis reveals a well-constrained region of interest for  $A_0$ and $k_s$, but we do not see a strong constraint coming from NANOGrav for the broadness parameter $k_e/k_s$. This lack of dependence is because NANOGrav only fits the ascending part of the second-order SGWB spectra, which does not depend very strongly on the width of the power spectrum. However, it is also evident that the NANOGrav data slightly disfavor the very narrow scalar power spectra compared to the broader ones.

\begin{figure}[t!]
    \centering
\includegraphics[width=0.49\linewidth]{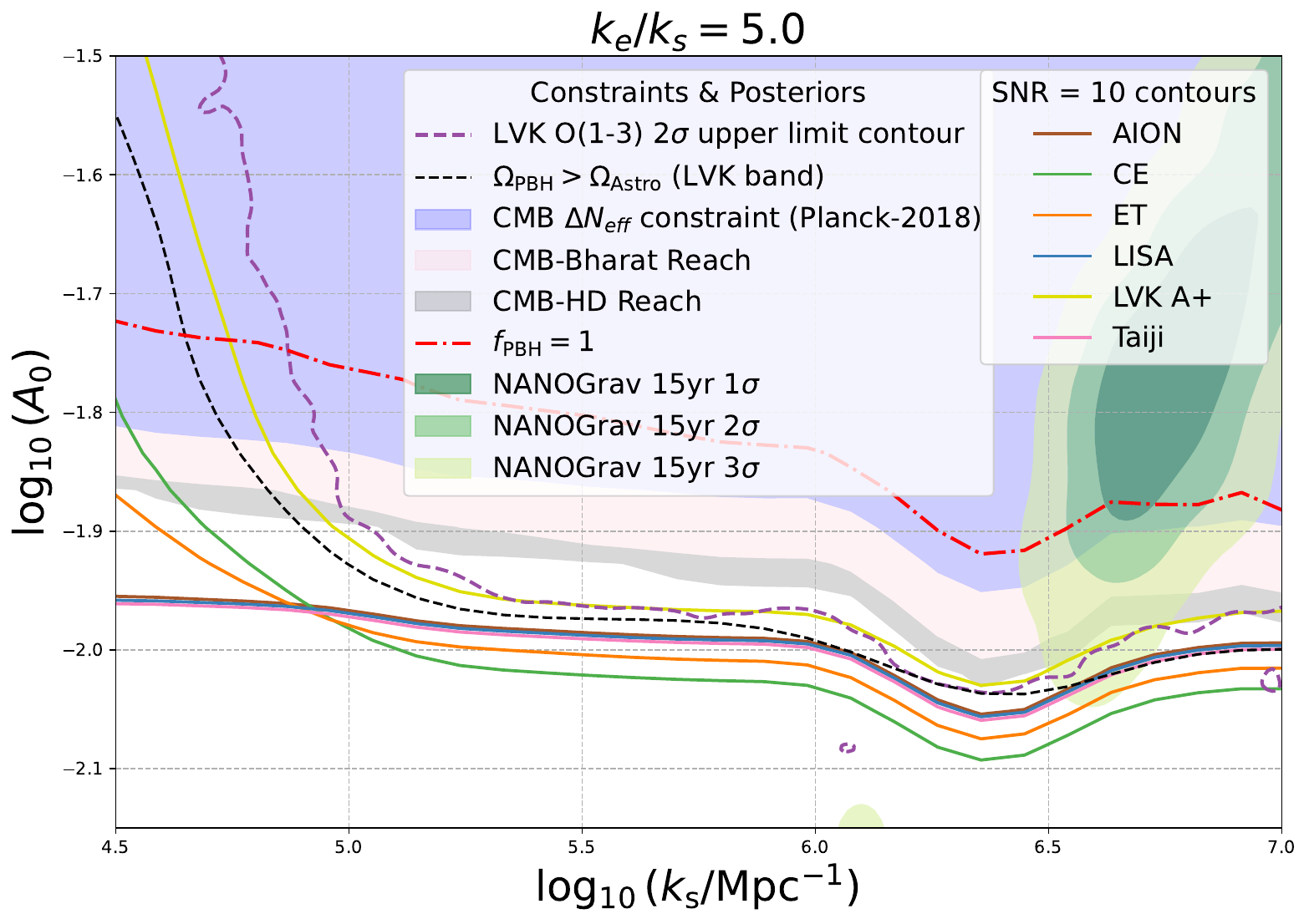}
\includegraphics[width=0.49\linewidth]{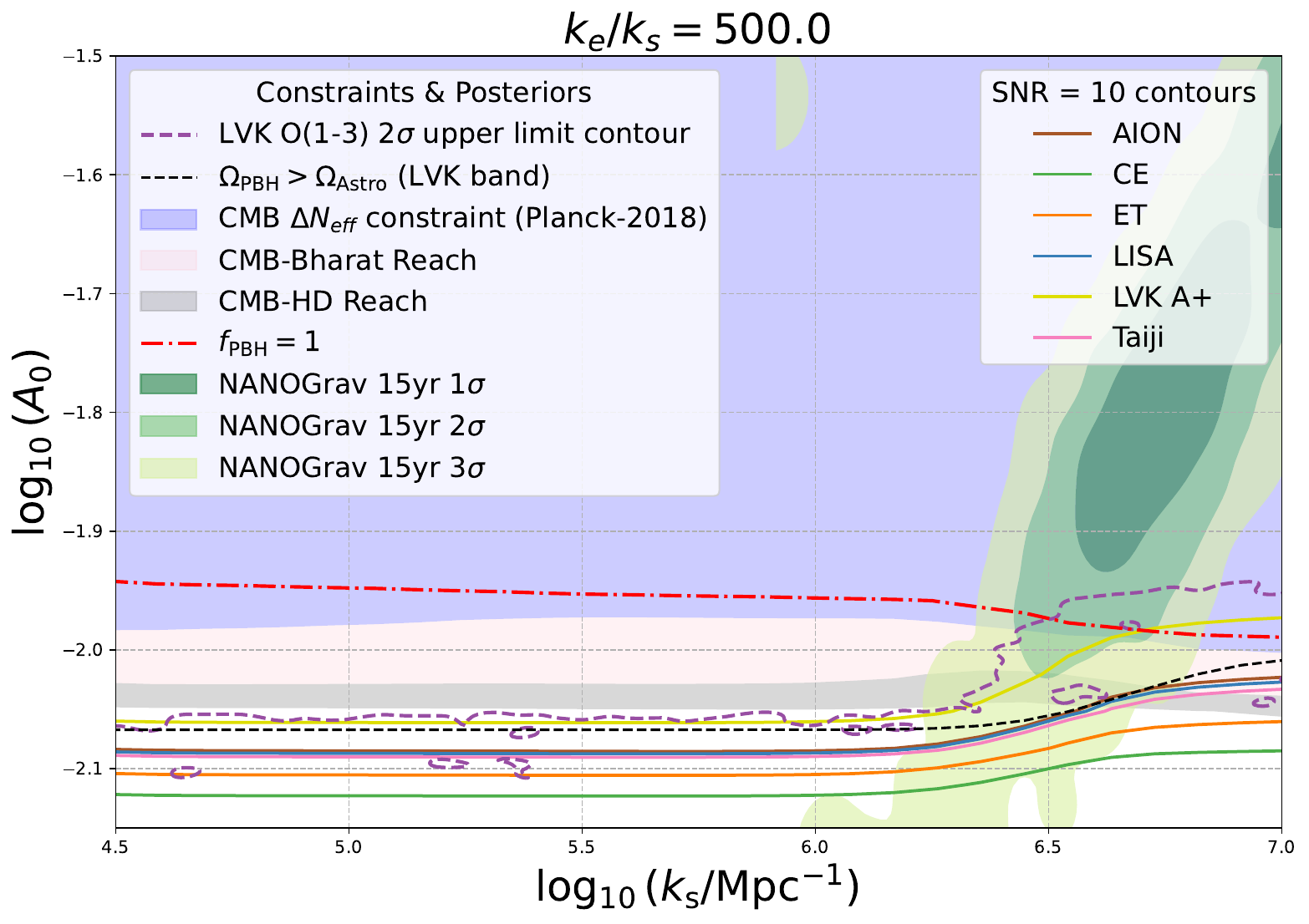}
\caption{Parameter space for the primordial scalar power spectrum, showing constraints from and detection prospects for the two distinct SGWBs. The \textbf{left panel} shows the results for a relatively narrow primordial spectrum with $k_e/k_s \approx 5 $, while the \textbf{right panel} shows the case for a much broader spectrum with $k_e/k_s \approx 500$. In both panels, the nested green shaded regions represent the $1\sigma$, $2\sigma$, and $3\sigma$ confidence levels where the second-order SGWB fits the NANOGrav 15-year data. The region above each solid line is the parameter space with $\text{SNR} > 10$ for various detectors with the SGWB produced by subsequent PBH mergers. The plot is overlaid with cosmological constraints, including the PBH dark matter limit ($f_{\text{PBH}}=1$, red dashed line) and the bound from excess radiation ($\Delta N_{\text{eff}}$) from Planck and projected sensitivities from future CMB missions like CMB-HD and CMB-Bharat \cite{Planck:2018jri, CMB-HD:2022bsz, CMBBharat:01}. The region below the black dashed line denotes the parameter space where the primordial SGWB spectra are overshadowed by the astrophysical SGWB originating from black hole black hole or neutron star black hole binaries in the LVK frequency band \cite{Regimbau:2011rp}.}
\label{fig:combined_parameter_space}
\end{figure}
\begin{figure}[t!]
    \centering
    \includegraphics[width=0.7\linewidth]{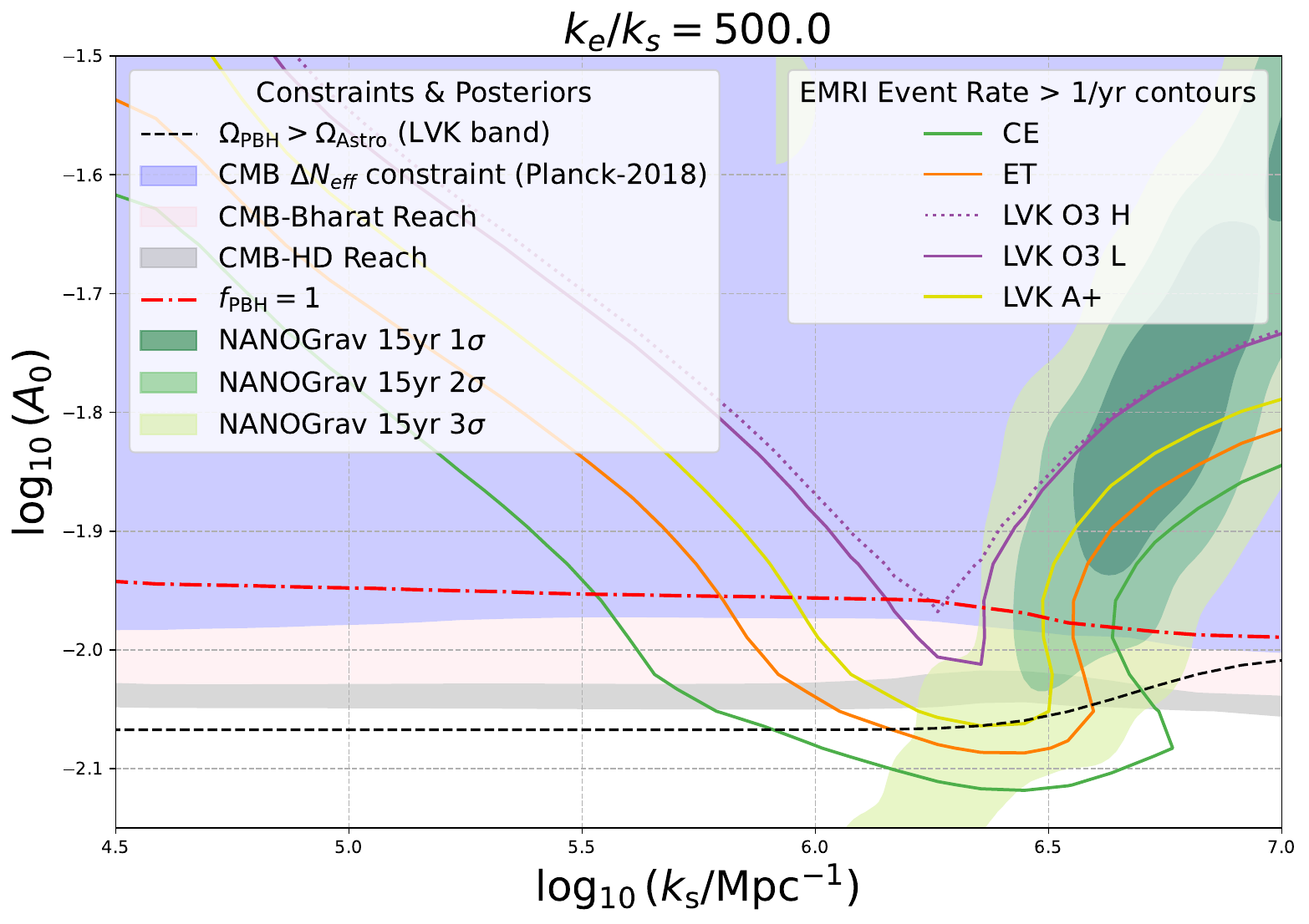}%final_emri_rate_contour_COMBINED_1000.0.pdf}
\caption{
    Combined parameter space for a primordial scalar spectrum, defined by its amplitude $A_0$ and wave number $k_s$. This plot highlights regions where PBHs would generate a detectable rate of mini-EMRI events. The solid colored lines are contours of one detectable event per year, representing the sensitivity of various detectors. The region above each line is the parameter space where the corresponding observatory would detect at least one EMRI event per year with a signal-to-noise ratio $\text{SNR} > 10$. The detectors shown are CE, ET, LVK O3-H, LVK O3-L and LVK A+  \cite{Abbott:2020wef, LIGO-P1200087, LIGO-T2000012, LIGO-T2000012-v2}. Overlaid are cosmological constraints and the $1\sigma$, $2\sigma$, and $3\sigma$ confidence levels from the NANOGrav 15-year dataset (green shaded regions).}
\label{fig:emri_parameter_space}
\end{figure}
\begin{figure}[t!]
    \centering
    \includegraphics[width=0.7\linewidth]{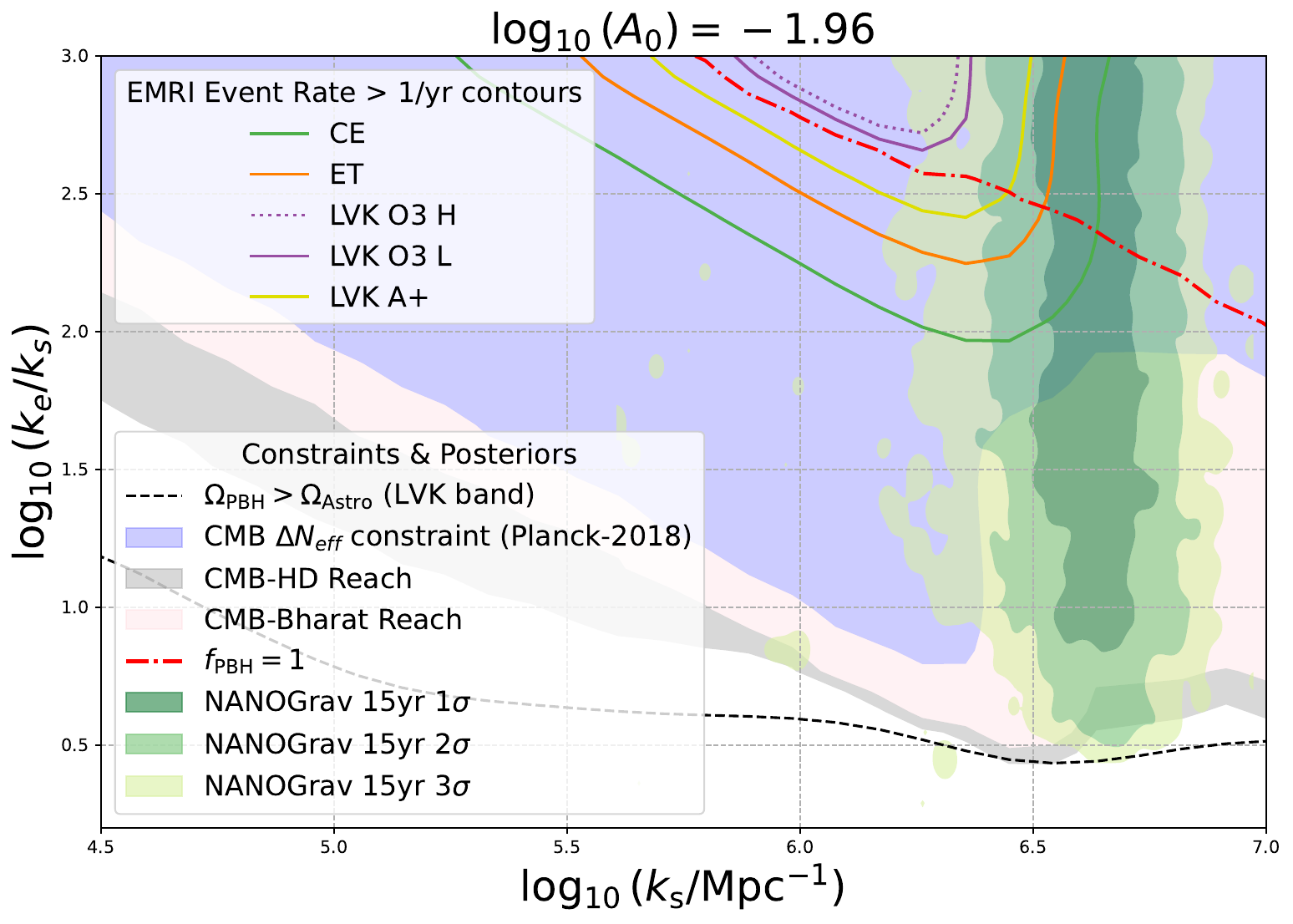}%final_emri_rate_contour_COMBINED_1000.0.pdf}
\caption{
   Here, we plot the mini-EMRI 1 event rate per year contours as a function of wave number $k_s$ and broadness parameter $k_e/k_s$, for a fixed amplitude $A_0$. The region above each line is the parameter space where the corresponding observatory would detect at least one EMRI event per year with a signal-to-noise ratio $\text{SNR} > 10$. This identifies the threshold of the broadness parameter below which mini-EMRI systems would not be detectable, along with the range of wave number $k_s$. On the other hand, two factors determine the confidence intervals for the peak wavenumber $k_s$: i) the frequency range for which GW detectors (in this case, the ground-based detectors like LVK, ET, and CE) are most sensitive, ii) the region of softer equation of state due to the QCD phase transition.}
\label{fig:emri_ks_krat}
\end{figure}

%\subsubsection{Constraints from LVK Searches}
In addition to comparing the second-order SGWB with PTA data, to constrain the SGWB from mergers, we also performed a separate parameter estimation for the data from the LVK O1-O3 observing runs~\cite{KAGRA:2021kbb}, with the results shown in Figure~\ref{fig:pygwb_corner}. Our analysis makes use of {\tt pygwb} ~\cite{Renzini:2023qtj, Renzini2024} following \cite{ PhysRevLett.118.121101, PhysRevD.100.061101, PhysRevD.104.022004, LIGOScientific:2025bgj}. This provides complementary constraints on the model parameters from an entirely different frequency band and GW source. The priors and posterior distributions of our LVK run are listed in Table~\ref{tab:lvk_param_summary}. This parameter search yields a log Bayes factor of $\log_{10}(\mathcal{B}) = -0.37 \pm 0.001$, indicating that the LVK data prefers Gaussian noise over the SGWB spectra from PBH mergers of our model. Thus, we can use the results of LVK analysis only to put upper bounds or constraints on our model parameters. The posterior distribution of $A_0$ in Figure~\ref{fig:pygwb_corner} clearly reflects this upper bound. It also shows the range of $k_s$ that are most sensitive to LVK data. On the other hand, the declining probability distribution with increasing values of the broadness parameter $k_e/k_s$ clearly indicates that the LVK data constrain broader PBH mass distributions more strongly than narrower ones.

 The question of how efficiently LVK data can constrain or detect the SGWB from the NANOGrav favored regions also depends on the broadness parameter. As we have already discussed, the ascending part of the second-order SGWB spectrum, which matches the NANOGrav data, has a dependence on the broadness of inflationary scalar spectra, and the merger-driven SGWB, which determines the LVK parameter space, also depends on the broadness of the PBH mass function. As we can see from Fig. \ref{fig:combined_parameter_space}, in both narrow and broad PBH mass function scenarios, the LVK O3 data excludes the $1\sigma$ region of the NANOGrav parameter space entirely. For a narrower spectrum, only a part of the $3\sigma$ region remains unconstrained; but for a broader spectrum, the LVK O3 data allows larger parts of the NANOGrav $2\sigma$ and $3\sigma$ regions.

\subsubsection{Prospects of future GW detectors and CMB observations}
In Fig.~\ref{fig:combined_parameter_space}, we vary $k_s$ and $A_0$ for fixed values of the broadness parameter $k_e/k_s$ and obtain SNR=10 contours of the corresponding SGWB from PBH mergers, for future ground and space-based GW detectors like LVK A+, CE, ET, LISA, AION, and Taiji. As SGWB increases with increasing values of $A_0$, for each of these SNR=10 contours shown in solid lines, the region above the line is detectable with SNR $> 10$. Along with the SNR contours, we also plot the constraints from the PBH overproduction bound with the red dashed line, NANOGrav contours in green, and LVK $2\sigma$ contour with a dashed violet line for the specific values of $k_e/k_s$. 

We also compare the SGWB from PBH mergers with the astrophysical SGWB~\cite{Regimbau:2011rp} in the frequency band of ground-based GW detectors. The region below the black dashed line represents the part of the parameter space where SGWB from ABH and NS binaries completely overshadows the SGWB from PBH binaries, thus making detection of the primordial SGWB in LVK, CE, or ET very difficult. However, the prospect of detecting primordial SGWB in this part of the parameter space, in future detectors like CE and ET, depends on how precisely these detectors can resolve and model the astrophysical binary black hole and neutron star population. The astrophysical SGWB is subject to significant uncertainty in modelling the compact binary coalescence (CBC) population. In this work, we use the $90\%$ credibility bound of the SGWB originating from different classes of astrophysical populations as obtained in the ref \cite{LIGOScientific:2025bgj}, which follows from earlier works \cite{Regimbau:2011rp, Callister:2020arv} with updated Compact Binaries catalogue \cite{LIGOScientific:2025pvj}.  To estimate the most conservative bound on the detectability of the primordial SGWB, we take the highest possible value of the astrophysical SGWB in this band as our reference value in any particular frequency and the region below the black dashed contours represents parameter space for which the value of $\Omega_{GW}$ from PBH models is smaller than the corresponding astrophysical background for all frequency points in the sensitivity band of ground-based detectors.

Along with GW detectors, the SGWB can also be constrained by the $\Delta N_{\text{eff}}$ bounds. The CMB observations and the requirement of consistency of Big Bang Nucleosynthesis (BBN) constrain the extra relativistic degrees of freedom in the form of the effective number of neutrinos $\Delta N_{\text{eff}}$ \cite{Steigman:1977kc, Kolb:1990vq, Dolgov:2002wy, Mukhanov:2005sc, Baumann:2022mni},
 \be
 \Delta N_{\rm eff} = \left\{\frac{8}{7}\left(\frac{4}{11}\right)^{-\frac{4}{3}}+N_{\rm eff}^{\rm SM}\right\} 
 \frac{\rho_{\rm SGWB }(\tau_{\rm EQ})}{\rho_{\rm rad}(\tau_{\rm EQ})}\, ,
\label{eqNeff2}
\ee
where we define $\Delta N_{\text{eff}}$ at matter radiation equality. This refers to the contribution coming from the SGWB component on top of the number of standard model neutrinos, $N_{\rm eff}^{\rm SM}  \approx 3.048$~\cite{Planck:2018vyg}. We show the constraints from Planck 2018 CMB data and the projected reach of the future of CMB detectors like CMB-HD and CMB-Bharat  \cite{Planck:2018jri, CMB-HD:2022bsz, CMBBharat:01} with lightly-shaded regions in Fig.~\ref{fig:combined_parameter_space}.
The difference between a broad and narrow inflationary power spectra, and thus a wide and narrow PBH mass distribution, is revealed by comparing the two panels of Figure~\ref{fig:combined_parameter_space}, which illustrates the impact of the primordial spectrum's width, $k_e/k_s$. While the NANOGrav posterior regions (driven by the formation-induced SGWB) show a weak dependence on the broadness parameter, the sensitivity contours for the merger-induced SGWB are strongly sensitive to the broadness parameter, as visible in Figure~\ref{fig:combined_parameter_space}. For a broader spectrum ($k_e/k_s \approx 500$, right panel), the contours shift downwards, indicating that detectors like ET and CE can probe much smaller primordial amplitudes ($A_0$) in the case of an extended PBH mass distribution. This is because a broader primordial spectrum generates a much larger total abundance of PBHs over an extended mass range, leading to a significantly enhanced merger rate and a stronger merger-induced SGWB. Consequently, for broad-spectrum scenarios, the merger-induced SGWB becomes an exceptionally powerful probe. Notably, in both cases, the entire 1$\sigma$ NANOGrav-favored regions are ruled out by the PBH overproduction bound.
In comparison, the future sensitivity of LVK, ET, CE, AION, LISA, and Taiji promises a robust, multi-band confirmation or refutation of the $2\sigma$ and $3 \sigma$ parameter region of NANOGrav. For a narrower spectrum ( $k_e/k_s \approx 5$ case), we can clearly see effects of the softer equation of state during QCD phase transition and the detector sensitivity band for ground-based detectors, while in the $k_e/k_s \approx 500$ case, these features get somewhat smoothened out due to the cumulative effect of a broader distribution, and we get nearly flat contours. It is essential to note that these contours depend on the specific formalism employed to estimate the PBH abundance, as the merger rate and SGWB from mergers depend strongly on the PBH abundance. Furthermore, it is well known that PBH abundance estimation varies significantly depending on the formalism used. However, we limit ourselves to the Peaks Theory formalism as outlined in section \ref{sec:framework}. 
\section{Mini-EMRIs as a Complementary Probe}
\label{sec:emri_probes}
While stochastic backgrounds probe the Universe on cosmological scales, mini-EMRIs provide a distinct window into the local population of PBHs. These systems are particularly valuable probes for three reasons. First, their higher gravitational-wave frequencies ($\sim$1--100 Hz) make them ideal targets for current and future ground-based observatories, unlike traditional EMRIs, which are primarily sources for LISA. Second, the detection of a population of EMRIs in this mass range would constitute a strong indication for an extended mass distribution of PBHs. Third, the SGWB signal from the merging PBH population is overshadowed by the SGWB of astrophysical origin, which comes solely from the mergers of ABH and NS populations, for the frequency range of ground-based detectors. This places the detection of individual events on an important footing as the only remaining channel to probe primordial GW sources. The following analysis quantifies the detection prospects for these sources.

\subsection{Methodology for Calculating Observable Event Rates}
To determine the number of detectable mini-EMRI events, we calculate the merger rate for a given population of compact objects, subject to a minimum signal-to-noise ratio (SNR) for a given gravitational-wave detector. The methodology involves computing the characteristic strain of the gravitational wave, evaluating the corresponding SNR, and integrating the merger-rate density over the sensitive volume defined by the SNR threshold.

\subsubsection{Signal-to-Noise Ratio}
The detectability of a gravitational-wave signal is quantified by the matched-filter signal-to-noise ratio (SNR), denoted by $\rho$. The square of the SNR is calculated by integrating the signal's power spectral density against the detector's noise power spectral density, $S_n(f)$, over the relevant frequency band \cite{Barack:2003fp} :
\begin{equation}
    \rho^2 = 4 \int_{f_{\text{min}}}^{f_{\text{max}}} \frac{|\tilde{h}(f)|^2}{S_n(f)} df,
\end{equation}
where $\tilde{h}(f)$ is the Fourier transform of the time-domain gravitational-wave strain, and $S_n(f)$ is the one-sided noise power spectral density of the detector. We only consider the waveform from the physical evolution of the binary system up to one year before its merger, starting from the frequency corresponding to the innermost stable circular orbit (ISCO), $f_{\rm ISCO}$, to the radiating frequency of the system one year earlier, $f_{\rm 1yr}$. 
The upper frequency cutoff, $f_{\text{ISCO}}$, is a function of the primary black hole's mass $M_1$ and dimensionless spin $a$~\cite{Bardeen:1972fi},
\begin{equation}
    f_{\text{ISCO}} = \frac{m}{2\pi} \frac{\tilde{\Omega}_{\text{ISCO}}}{M_1} \, ,
\end{equation}
where m is the harmonic mode, the dimensionless orbital angular velocity is defined as $\tilde{\Omega} \equiv M_1 \Omega = 1/(\tilde{r}^{3/2}+a)$ and $\tilde{r} \equiv r/M_1$ with $r$ the Boyer–Lindquist radial coordinate of the orbit. The dimensionless radius $\tilde{r}_{\text{ISCO}}$ would include a correction factor that depends on the black hole's spin through the relations:
\begin{align}
    Z_1 &= 1 + (1-a^2)^{1/3} \left[ (1+a)^{1/3} + (1-a)^{1/3} \right], \\
    Z_2 &= \sqrt{3a^2 + Z_1^2}, \\
    \tilde{r}_{\text{ISCO}} &= 3 + Z_2 - \text{sgn}(a) \sqrt{(3-Z_1)(3+Z_1+2Z_2)}.
\end{align}
For the non-spinning Schwarzschild case ($a=0$) that we consider, this simplifies to $\tilde{r}_{\text{ISCO}}=6$, giving $f_{\text{ISCO}} \approx 4.4\,\text{kHz}\,(M_\odot/M_1)$~\cite{Guo:2019sns}. The lower frequency $f_{\text{1yr}}$, is found by inverting the formula for the time remaining until merger, $T(f)$ ~\cite{Finn:2000sy},
\begin{equation}
    T(f) = \frac{5}{256} \frac{M_1}{\eta \tilde{\Omega}^{8/3}} \mathcal{T}\, .
    \label{eq:time_to_merger}
\end{equation}
Here, we use the mass ratio $\eta \equiv M_2/ M_1$ and $\mathcal{T}$ accounts for relativistic corrections. Thus, the integration limits, $f_{\text{min}}$ and $f_{\text{max}}$, are set by the overlap between the detector's sensitivity band and the frequency range from  $f_{\rm 1yr}$ to $f_{\rm ISCO}$. For convenience, we express the SNR in terms of the characteristic strain, $h_c(f) = 2f|\tilde{h}(f)|$:
\begin{equation}
    \rho^2 = \int_{f_{\text{min}}}^{f_{\text{max}}} \frac{h_c(f)^2}{f^2 S_n(f)} df.
\label{eq:snr_hc}
\end{equation}

\subsubsection{Characteristic Strain Waveform}
We consider a quasi-circular orbit in this work for simplicity and follow Eq. (3.14b) of ~\cite{Finn:2000sy}, which also takes into account the relativistic correction factors based on the Teukolsky-Sasaki-Nakamura formalism~\cite{Teukolsky:1973ha, Sasaki:1981sx} for the characteristic strain $h_{c,m}$ expression,
\begin{align}
    h_{c,m} &=\mathcal{A}_m \frac{\eta^{1/2}M_2}{D_L} \times \tilde{\Omega}^{(2m-5)/6} \mathcal{H}_{c,m}, \quad m \ge 2\ .
    \label{eq:CharStrain}
\end{align}
Here, $\mathcal{A}_m$ is a dimensionless prefactor for a given harmonic mode $m$, derived from the quadrupole radiation formula \cite{Finn:2000sy},
\begin{align}
  \mathcal{A}_m &= \sqrt{\frac{5(m + 1)(m + 2)(2m + 1)!m^{2m}}{12\pi (m - 1)![2^m m! (2m + 1)!!]^2}}.
\end{align}
The $m=2$ harmonic dominates the gravitational waveform. We estimate $h_c$ only for the dominant $m=2$ mode. The dynamics of the binary interaction are captured by the mass ratio $\eta$, and $h_c$ is inversely proportional to $D_L$, the distance to the Earth. Finally, $\mathcal{H}_{c,2}$ is the dimensionless relativistic correction factor, which accounts for general relativistic effects near the central object. The numerical values of this factor were tabulated in Ref.~\cite{Finn:2000sy}. Since we confine ourselves to PBH–PBH binary systems, we can assume non-spinning Schwarzschild black holes. For non-spinning PBHs with circular orbits where we can take the adiabatic approximation, it is evident from Fig. \ref{fig:all_comparisons} that  Eq. \eqref{eq:CharStrain} is in excellent agreement with the outputs of updated numerical packages like 
\texttt{FastEMRIWaveforms}~\cite{Chua:2020stf, Katz:2021yft, Speri:2023jte, Chapman-Bird:2025xtd}.

\subsubsection{Observable Event Rate Calculation}
To calculate the total number of observable merger events per year, $N_{\text{obs}}$, we follow a multi-step process.

\paragraph{1. Maximum Observable Distance:}
We first establish a detection threshold of $\rho_{\text{thresh}} = 10$. Since the characteristic strain is inversely proportional to the luminosity distance ($h_c \propto 1/D_L$), the SNR also follows this relation ($\rho \propto 1/D_L$). For each pair of masses $(m_i, m_j)$, we can thus determine the maximum luminosity distance, $D_{L, \text{max}}$, at which the system would be detectable:
\begin{equation}
    D_{L, \text{max}}(m_i, m_j) = D_{L, \text{ref}} \cdot \frac{\rho(D_{L, \text{ref}})}{\rho_{\text{thresh}}},
\end{equation}
where we calculate a reference SNR, $\rho(D_{L, \text{ref}})$, at a fixed distance $D_{L, \text{ref}}$ (e.g., 1 Gpc).

\paragraph{2. Sensitive Volume and Number of Mergers in a Given Mass Interval:}
The maximum observable distance $D_{L, \text{max}}$ defines a sensitive volume for each binary configuration. We also account for the evolution of the merger rate with redshift by including a cosmic evolution factor, calculated from the relationship between cosmic time and redshift, $t(z)$. For a given $D_{L}$, we first find the corresponding redshift $z$, which allows us to estimate the corresponding evolution factor and define the detectable merger rate in the volume $D_{L, \text{max}}$ as
\begin{equation}
\mathcal{E}(m_i,m_j)=\int_{0}^{D_{L, \text{max}}} \frac{dR_{E2}}{d m_1 d m_2}\Big\vert_{t=t(z)} 4 \pi D_{L}^2 \, dD_{L}.
\end{equation}

\paragraph{3. Total Event Rate:}
Finally, the total observable event rate is obtained by integrating $\mathcal{E}(m_i,m_j)$ over all possible mass combinations, weighted by their corresponding sensitive volumes and evolution factors:
\begin{equation}
    N_{\text{obs}} = \int_{m_i/m_j < 10^{-4}} \mathcal{E}(m_i,m_j) \, dm_i \, dm_j.
\label{eq:n_obs}
\end{equation}
This integral is computed numerically over the grid of masses for which the merger-rate density is defined, subject to the EMRI mass-ratio constraint.

\subsection{Results: EMRI Event Rates and Detection Prospects}
 The input for the event-rate calculation is the differential merger-rate density, shown in Figure~\ref{fig:merger_rate_contours}. This plot illustrates how the landscape of PBH mergers is a direct consequence of the underlying primordial power spectrum. For instance, the model with a very broad spectrum ($k_e/k_s = 500$, black contours) produces a merger rate that is orders of magnitude higher than the others and is heavily distributed in the different mass regions. In contrast, the model with a narrower spectrum ($k_e/k_s = 10$, green contours) predicts a localized distribution with much smaller amplitude, with higher merger rates for binaries near the solar-mass scale, reflecting the peaked nature of its corresponding PBH mass function. 

Here, the characteristic mass of merging PBHs acts as an interesting tracer of the inflationary physics. By integrating these merger rates over the detector-sensitive volumes, we can map the detection prospects of the PBH parameter space, which is the key result of this analysis, shown in Figure~\ref{fig:emri_parameter_space}. This plot presents the prospect of using resolvable mini-EMRI events as an independent test of the scenario in which the broader mass distribution of PBHs can be a candidate to explain the NANOGrav signal. The regions above the solid and dotted (in the case of LVK O3 H) colored lines represent the parameter space where future observatories would detect at least one mini-EMRI event per year with an SNR greater than 10. 

Crucially, the parameter space favored by the NANOGrav 15-year data (green shaded regions) shows significant overlap with the sensitivity reach of next-generation detectors. For example, a large portion of the  $2\sigma$ and $3\sigma$ NANOGrav regions that are not ruled out by the overproduction of PBHs (red dashed line in Figure~\ref{fig:emri_parameter_space}) predicts a detection rate of more than one event per year in the LVK A+, ET, and CE. It is also important to note that in the LVK frequency band, part of this parameter space is obscured by the astrophysical SGWB background originating from ABH and NS mergers~\cite{Regimbau:2011rp} (dashed black line in Figure~\ref{fig:emri_parameter_space}), and is therefore inaccessible to detection through the SGWB channel with any future LVK detectors, CE, or ET.

This implies that if the NANOGrav signal indeed arises from a broad primordial scalar spectrum, individual mini-EMRI events would provide a unique opportunity to probe such a scenario with future detectors, such as LVK A+, CE, and ET. The detection of both the stochastic background by PTAs and a corresponding population of resolvable mini-EMRIs by ET or CE would provide a powerful, multi-messenger confirmation of this entire framework or ruling out of the scenario.

The option of probing the PBH population with individual mini-EMRI systems only comes for broader mass distribution of PBHs, as shown in Fig.~\ref{fig:emri_ks_krat}, where we fix a specific value for $A_0$ and vary both $k_s$ and the broadness parameter $k_e/k_s$. It is evident from Fig.~\ref{fig:emri_ks_krat} that the detectable mini-EMRI systems are achieved in only the case of sufficiently high values of $k_e/k_s$, or broader PBH mass distribution. On the other hand, the nearly vertical distribution of the NANOGrav contours reaffirms that, for fixed values of $A_0$, the NANOGrav data is not very sensitive to variations in $k_e/k_s$, while it shows a very localized preference for the wavenumber of the PBH peak, $k_s$.

\section{Discussion and Conclusions}
\label{sec:conclusions}

In this work, we have presented a comprehensive multi-messenger analysis of PBHs formed from a broad inflationary power spectrum, which leads to the formation of PBHs around the epoch of the QCD phase transition. The change in the relativistic degrees of freedom during the QCD phase transition results in a softening of the Universe's equation of state, leading to more efficient PBH formation. Our analysis connects the primordial spectrum to three distinct gravitational-wave signatures: two different channels of SGWB, along with the detectable individual mini-EMRI systems. Mini-EMRI systems are of particular interest in the case of a broader PBH mass distribution and play a crucial role in probing the extended PBH mass fraction. We also perform Bayesian analysis of the NANOGrav 15-year data release~\cite{NANOGrav:2023hvm} for second-order SGWB, with \texttt{PTArcade}~\cite{Mitridate:2023oar} and LVK O1-O3 data~\cite{KAGRA:2021kbb, LIGO-G2001287} for SGWB from PBH mergers using the {\tt pygwb}~\cite{Renzini:2023qtj, Renzini2024} and~\texttt{BILBY}~\cite{Ashton:2018jfp}, along with parameter space scan for future ground-based and space-based GW detectors. Here we list some of the interesting points we find,
\begin{itemize}
    \item As shown in Fig. \ref{mass}, we use a flat amplification of the inflationary scalar power spectrum to obtain a broadly distributed PBH mass function, which is a requirement to form mini-EMRI systems. While standard EMRI systems with supermassive primary PBH would be a subject of strong constraints due to CMB accretion and $\mu$-distortion bounds from FIRAS~\cite{1994ApJ...420..439M, 2012ApJ...758...76C, 2014PhRvL.113f1301J, Carr:2020gox}, the constituent PBHs in the case of mini-EMRI will be relatively less restricted. Still, it can be constrained by micro-lensing observations, such as those from EROS~\cite{EROS-2:2006ryy, Carr:2017jsz}. As these constraints will also depend on the broadness of the PBH mass distribution, in this study, we do not explicitly use these constraints.

    \item Our results strongly depend on the rate of PBH formation and mergers. We discuss our adopted formalism regarding this in section \ref{PBH-merger}, and plot the merger rate in logarithmic mass intervals for three sets of parameters (same as Fig.~\ref{mass}) in Fig.~\ref{fig:merger_rate_contours}. However, the merger rate calculation formalism becomes less accurate as we approach broader mass distributions. We leave further study in this direction for the future.

    \item Our scenario with a broad PBH mass function naturally leads to two different channels of SGWB. The first is the second-order SGWB originating from the PBH formation, just after their horizon entry. As shown in Fig. \ref{omega}, this component falls within the observation band of PTA observatories, such as NANOGrav. On the other hand, another source of the SGWB is the mergers from the formed binary populations, which contain both Intermediate mass ratio (IMR) and mini-EMRI systems. We use different numerical waveforms for these different types of systems as illustrated in Fig. \ref{fig:all_comparisons}. The peak of these merger-induced SGWB is typically detectable in the ground-based GW observatories, and the tail part can be detected in the space-based GW observatories, as we can infer by comparing with the power-law integrated sensitivity curves of different detectors shown in Fig. \ref{omega}.
    
    \item  Our Bayesian analysis with NANOGrav 15-year data identifies the $1 \sigma$, $2 \sigma$, and $3 \sigma$ regions of the parameter space of the inflationary power spectra, as shown in Figure~\ref{nanograv} and Table~\ref{pta_params_3p_full}. The merger-induced SGWB is independently constrained by our Bayesian analysis of LVK O1-O3 data, which probes a different frequency band. Furthermore, our detailed parameter space scan for the detectability (SNR $>$ 10) of this SGWB in future observatories (Figure~\ref{fig:combined_parameter_space}) reveals that while the $1\sigma$ posterior regions of NANOGrav are in tension with current LVK bounds~\cite{LIGOScientific:2021nrg}, the $2\sigma$ and $3\sigma$ regions remain viable and present a clear target for next-generation detectors.

    \item  The ability of LVK data to constrain or detect the SGWB originating from the NANOGrav-favored regions is also influenced by the broadness parameter. As discussed earlier, both the fit to the NANOGrav signal and the LVK constraints are sensitive to the width of the inflationary scalar power spectrum, and thus the broadness of the PBH mass function. As shown in Fig. \ref{fig:combined_parameter_space}, LVK O3 observations rule out the entire $1\sigma$ region of the NANOGrav parameter space for both narrow and broad PBH mass functions. In the narrow case, only a small portion of the $3\sigma$ region remains allowed; whereas in the broader spectrum case, LVK O3 data permits larger fractions of the NANOGrav $2\sigma$ and $3\sigma$ regions.

    \item In Fig. \ref{fig:combined_parameter_space}, we show the parameter space with detectable merger-driven SGWB spectra for the other detectors, along with the NANOGrav and LVK contours, the PBH overproduction bound, and the detection aspects of $\Delta_{N_{\rm eff}}$ in present and future CMB observations. The comparison of two panels of Fig. \ref{fig:combined_parameter_space} also illustrates the clear downward shift of detectable parameter space from a narrowly peaked spectrum to a broader inflationary scalar spectrum.

    \item The third channel of observation that we study in this work involves the detection of the gravitational waves generated from the individual mini-EMRI systems. This channel is only available for a broad enough PBH mass distribution, as shown in Fig. \ref{fig:emri_ks_krat}.

    \item In Fig.~\ref{fig:emri_parameter_space}, we show the prospects of at least one detectable (SNR $>$ 10) mini-EMRI system per year in different future ground-based detectors. Particularly due to the masking of primordial SGWB by the astrophysical SGWB~\cite{Regimbau:2011rp} in the frequency band of ground-based detectors, the SGWB channel becomes relatively ineffective in probing a large part of the parameter space (shown as the region below the black dashed line in Fig~\ref{fig:combined_parameter_space}, Fig~\ref{fig:emri_parameter_space}, and Fig.~\ref{fig:emri_ks_krat}). This leaves the detection of mini-EMRI systems as the only remaining probe. Thus, to constrain or detect the population of PBHs with a broader mass distribution (characterized by the broadness parameter $k_e/k_s$), detection of mini EMRI systems becomes a more effective channel than the stochastic background.  
\end{itemize}

\noindent Taken together, our findings support a scenario in which broadly peaked, amplified scalar perturbations can lead to an extended PBH mass distribution and simultaneously explain the nHz second-order SGWB in NANOGrav, while producing merger signals in the Hz to kHz band. The prospect of detecting both the stochastic backgrounds and individual mini-EMRI events provides a powerful observational pathway to verify or constrain the broader PBH distribution scenario and distinguish between astrophysical and primordial origins of black hole populations. 

It is interesting to note that, for the mini-EMRI detection part, we primarily focus on the waveform from the inspiral part and take into account the systems that merge within the lifetime of our universe; however, there can be other mini-EMRI systems that are also radiating and will merge in the future, but we are not taking those systems into account. A more detailed study of PBH binary formation can help identify these systems, which we leave for future work. In this work, we have not taken into account the effects of clustering of PBHs. As we deal with extended mass distributions with lighter PBHs, this effect can be significant~\cite{Crescimbeni:2025ywm}. In future work, we plan to develop a more accurate estimate of PBH merger rates that is appropriate for the extended mass distribution of PBHs, also taking into account the effects of three-body encounters and clustering.
\section*{Note added}
During the completion of this work, another preprint appeared on
arXiv~\cite{Gouttenoire:2025jxe}, which also studies PBHs forming around the epoch of QCD phase transition and their observable imprints in SGWB, with the NANOGrav and LVK data, for a log-normal peak in inflationary scalar power spectrum. While ref.~\cite{Gouttenoire:2025jxe} also takes into account the effects of local non-Gaussianity, the focus of our work remains on identifying the broader mass distribution of PBHs with SGWB as well as the detection of mini-EMRI systems, which we find to be more effective in distinguishing the primordial signal from the gravitational waves generated by the astrophysical sources like NS and ABH binaries.

\section*{Acknowledgements}

The authors thank Tore Boybeyi for very fruitful discussions and feedback on the draft. The authors thank Marek Lewicki for the AION-km sensitivity data. The authors also thank Andrew Miller, Juan García-Bellido, and Sébastien Clesse for interesting discussions during the presentation of the preliminary version of the results in February. NB thanks Mai Qiao and Alba Romero-Rodríguez for interesting discussions on the analysis of LVK data with \texttt{pygwb}. NB wishes to thank his parents and family members for their unwavering support during the completetion of this work. This work was supported by the International Center for Theoretical Physics Asia Pacific and the University of Chinese Academy of Sciences. This work was also supported by the National Natural Science Foundation of China (NSFC) under Grant No. 125471042 and under Grant No. 12347103.
    
%\input{appp}
% --- Bibliography ---
\bibliographystyle{JHEP}
\bibliography{mybib}

\end{document}